\documentclass{emulateapj}

\usepackage{psfig}
\usepackage{amsmath}
\usepackage{amssymb}
\usepackage{graphicx}
\usepackage{appendix}
\usepackage{color}

\newcommand{\lSect}[1]{{\label{sec:#1}}}
\newcommand{\lFig}[1]{{\label{fig:#1}}}
\newcommand{\lEq}[1]{{\label{eq:#1}}}
\newcommand{\lTab}[1]{{\label{tab:#1}}}
\def\gtaprx {\lower .1ex\hbox{\rlap{\raise .6ex\hbox{\hskip .3ex
	{\ifmmode{\scriptscriptstyle >}\else
		{$\scriptscriptstyle >$}\fi}}}
	\kern -.4ex{\ifmmode{\scriptscriptstyle \sim}\else
		{$\scriptscriptstyle\sim$}\fi}}}
\def\ltaprx {\lower .1ex\hbox{\rlap{\raise .6ex\hbox{\hskip .3ex
	{\ifmmode{\scriptscriptstyle <}\else
		{$\scriptscriptstyle <$}\fi}}}
	\kern -.4ex{\ifmmode{\scriptscriptstyle \sim}\else
		{$\scriptscriptstyle\sim$}\fi}}}
\newcommand{\FIGFF}[2]{{\ref{fig:#2}{#1}}}

\newcommand{\FIG}[2]{{Fig.~\FIGFF{#1}{#2}}}
\newcommand{\Fig}[1]{{\FIG{}{#1}}}

\newcommand{\Sectff}[1]{{\ref{sec:#1}}}
\newcommand{\Sect}[1]{{\S~\Sectff{#1}}}

\newcommand{\Eqref}[1]{{\ref{eq:#1}}}
\newcommand{\Eqff}[1]{{(\Eqref{#1})}}
\newcommand{\eqff}[1]{{\Eqref{#1}}}

\newcommand{\Eq}[1]{{eq.~\Eqff{#1}}}
\newcommand{\eq}[1]{{equation~\eqff{#1}}}

\newcommand{\Msun}{\ensuremath{\mathrm{M}_\odot}}
\newcommand{\Tab}[1]{{Table \ref{tab:#1}}}

\begin{document}

%\shorttitle{Evolution of 7 - 11 \ \Msun Stars}
%\shortauthors{Woosley and Heger}

\title{The Remarkable Deaths of 9 - 11 Solar Mass Stars}

\author{S. E. Woosley\altaffilmark{1} and Alexander Heger\altaffilmark{2}}

\altaffiltext{1}{Department of Astronomy and Astrophysics, University
  of California, Santa Cruz, CA 95064; woosley@ucolick.org}

\altaffiltext{2}{Monash Center for Astrophysics, Monash University,
  VIC 3800, Australia, alexander.heger@monash.edu}

\begin{abstract}
The post-helium burning evolution of stars from $7\,\Msun$ to
$11\,\Msun$ is complicated by the lingering effects of degeneracy and
off-center ignition.  Here stars in this mass range are studied using
a standard set of stellar physics.  Two important aspects of the study
are the direct coupling of a reaction network of roughly $220$ nuclei
to the structure calculation at all stages and the use of a subgrid
model to describe the convective bounded flame that develops during
neon and oxygen burning. Below $9.0\, \Msun$, degenerate oxygen-neon
cores form that may become either white dwarfs or electron-capture
supernovae.  Above $10.3\,\Msun$ the evolution proceeds ``normally''
to iron-core collapse, without composition inversions or degenerate
flashes. Emphasis here is upon the stars in between which typically
ignite oxygen burning off center. After oxygen burns in a convectively
bounded flame, silicon burning ignites in a degenerate flash that
commences closer to the stellar center and with increasing violence
for stars of larger mass. In some cases the silicon flash is so
violent that it could lead to the early ejection of the hydrogen
envelope.  This might have interesting observable consequences.  For
example, the death of a $10.0\,\Msun$ star could produce two
supernova-like displays, a faint low energy event due to the silicon
flash, and an unusually bright supernova many months later as the low
energy ejecta from core collapse collides with the previously ejected
envelope.  The potential relation to the Crab supernova is discussed.
\end{abstract}


\section{INTRODUCTION}
\lSect{intro}

The post-carbon burning evolution of stars near the transition from
those that leave white dwarfs to those that make supernovae has always
been challenging to model \citep{Bar74}.  For non-rotating
stars of solar composition, the most interesting mass range is from
$7\,\Msun$ to $11\,\Msun$.  For other metallicities, the range
may vary \citep{IH13}.  In these stars, the effects of degeneracy
persist and, combined with neutrino cooling by the plasma- and
pair-processes, lead to temperature inversions, off-center shell
ignition, and composition inversions.  Despite these computational
challenges, the late stages of evolution of these stars have been
studied extensively.  Representative historical and important
  recent calculations have been done by
\citet{Miy80,Nom84,Hil84,Nom87,May88,Dom93,Tim94,Gut96,Gar97,
  Ibe97,Rit99,Sie06,Poe08,Jon13,Tau13,Tak13,Doh14,Jon14}, and
\citet{Mor14}.

An obscure, but relevant historical reference is \citet{Woo80}.  This
was the first, and so far as we know, only study to report a quite
different outcome for the explosion of $10\,\Msun$ stars - a violent
flash at the onset of silicon ignition that ejected the envelope of
the star.  After the envelope was ejected, the remaining bound core,
consisting of a mixture of silicon and iron, completed its silicon
burning evolution and collapsed, about a year later, to a neutron
star.  The collision between the matter ejected when the iron core
collapsed and the previously ejected envelope created an unusually
bright supernova.  For a time, an informal, unpublished discussion
went on between the Japanese \citep[e.g.][]{Miy80,Nom84,Miy87}
and US groups as to the actual fate of $10\,\Msun$ stars - electron
capture in a core of neon and oxygen resulting in collapse or silicon
deflagration followed by iron-core collapse, with the ultimate
agreement being that probably both happened for some narrow range of
masses \citep[e.g.][]{Woo86}.

Here we return to this issue with the same code used in 1980, but with
improved stellar and nuclear physics. We find, as anticipated, that
both electron-capture supernovae and silicon deflagration are common
outcomes for stars near $10\,\Msun$.  For the assumed physics, the
production of a neon-oxygen core that collapses due to electron
capture on unburned fuel happens for solar metallicity stars
  below $9.0\,\Msun$, while degenerate silicon flashes characterize stars
  from $9.0\,\Msun$ to $10.3 \Msun$. In the range $9.8\,\Msun$ to
  $10.3\,\Msun$, the flash is particularly violent and could lead to
  envelope ejection prior to iron core collapse.

In \Sect{physics} the stellar and nuclear physics used in the models
is discussed. In \Sect{presn} the evolution below $9.0\,\Msun$ is
briefly reviewed.  This subject is worth revisiting, if only to
  set some fiducial masses for the KEPLER code. We find, as have many
others, regions of mass where carbon-oxygen (CO) dwarfs are the
outcome and others where neon-oxygen-magnesium (ONe) white dwarfs
result. If such stars retain their hydrogen envelope until death,
thermonuclear supernovae (CO-dwarfs) or electron-capture supernovae
(ONe dwarfs) will result. In \Sect{flame} and \Sect{sishell}, we
  discuss the stars that, following the propagation of convectively
  bounded oxygen and silicon burning flames, produce iron cores that
  collapse to neutron stars. A novel treatment of oxygen burning
  flames (oxygen CBFs) is employed that incorporates a subgrid model
  for the flame propagation as a function of temperature in a full star
  calculation of the evolution (\Sect{oflame}). The silicon flash and
  its prompt effects are studied for each model where it occurs. Since
  the flash frequently ignites off-center, one must also follow the
  subsequent propagation of a {\sl silicon} burning CBF to the center
  after the core contracts and reignites burning (\Sect{siflame}). It
  is found that the Rayleigh-Taylor instability plays a major role in
  the propagation of this silicon-burning flame.

In \Sect{siflash}, we discuss, in greater detail, those stars where
silicon ignites explosively, calculating the light curves for a few
cases where a large amount of silicon burns (\Sect{ultra}).  Some of
these events can be brighter than a Type Ia supernova (SN Ia) for a
month or two.  Finally, we conclude (\Sect{conclude}) with some
speculations regarding the nature of the supernova that made the Crab
Nebula and speculate as to how our results might change in the case of
a more realistic multi-dimensional simulation.

\section{PHYSICS EMPLOYED}
\lSect{physics}

The full evolution of stars with main sequence masses in the range
$6.5\,\Msun$ to $13.5\,\Msun$ was calculated (\Tab{endstate}) using
the KEPLER code \citep{Wea78,Woo02}.  All stars had an initially solar
composition and, with one exception (\Sect{rotate}), were not
rotating.  The solar abundance set employed was from \citet{Lod03}
with $X = 0.711$, $Y = 0.274$, and metallicity, $Z = 0.015$.  Mass
loss was included using standard prescriptions and the nuclear
reaction rates and opacities were the same as used in many previous
studies \citep[e.g.,][]{Woo02,Woo07a}.  The grid of stellar masses
calculated was non-uniform and focused upon stars where the nature of
silicon burning was rapidly varying.  For stars lighter than
$9.0\,\Msun$, the ``end state'' was not determined since the
calculations were halted once a very thin helium shell had formed.
These stars would evolve as super-asymptotic giant branch (SAGB) stars
and may produce electron-capture supernovae or just end their lives as
ONe white dwarfs \citep[e.g.,][]{Poe08,Jon13,Doh14}.

An important aspect of the calculation was the use of a
moderate-sized nuclear reaction network directly coupled to the
stellar structure \citep[for prior examples,
    see][]{Woo04,Jon13}.  The network at the presupernova stage
typically contained approximately 230 isotopes from carbon through
germanium.  A representative network included $^{1-2}$H, $^{3-4}$He,
$^{6-8}$Li, $^{7,9-11}$Be, $^{8,10-14}$B, $^{11-14}$C, $^{13-15}$N,
$^{14-18}$O, $^{16-19}$F, $^{19-23}$Ne, $^{21-25}$Na, $^{23-28}$Mg,
$^{25-29}$Al, $^{27-34}$Si, $^{30-35}$P, $^{31-38}$S, $^{34-39}$Cl,
$^{36-42}$Ar, $^{37-43}$K, $^{40-48}$Ca, $^{41-49}$Sc, $^{44-52}$Ti,
$^{45-53}$V, $^{48-56}$Cr, $^{50-57}$Mn, $^{52-60}$Fe, $^{54-61}$Co,
$^{56-64}$Ni $^{57-65}$Cu, $^{60-66}$Zn, $^{64-80}$Ga $^{64-82}$Ge and
neutrons. A test case (10.5 \Msun) that used a larger network of 365
isotopes, the difference being more neutron-rich isotopes for the same
elements above calcium, gave nearly identical results for the
presupernova composition and structure. For example, the silicon core
masses differed by less than 0.005 \Msun, and the central values of
$Y_e$ for the two cases were 0.4334 and 0.4324 for the two
cases. Heavier elements and more isotopes could easily have been
included, and would be necessary to study the s-process in these
  stars, but the purpose here was to survey the structure of
presupernova stars, not their heavy element nucleosynthesis.  It was
important to link this network directly to the structure during
post-helium burning evolution because of the considerable electron
capture that goes on in the degenerate stellar cores prior to silicon
ignition \citep{Jon13}.  It was also important to include the
proper nuclear physics for silicon burning itself when the composition
consists chiefly of $^{30}$Si and $^{34}$S, not $^{28}$Si and
$^{32}$S. An ``alpha-network'' would have been wholly inadequate and
the ``QSE'' network normally used for silicon burning in KEPLER would
not treat neutron-rich silicon burning very accurately (even though it
does include special reactions for the destruction of $^{30}$Si).

The network was ``adaptive'' \citep{Rau02}. Any isotope that
would have appreciable abundance was automatically and dynamically
added as needed during the calculation. Up until the time of iron core
collapse, the electron fraction, $Y_\mathrm{e}$, within the helium
core stayed in the range $0.43$ to $0.50$. Thus the composition and
weak interactions were well represented by the moderate-sized network.

Weak interactions during oxygen and silicon burning affect the
evolution, mostly by altering the Chandrasekhar mass.  Here, as in
past studies \citep{Woo07a,Heg01}, the ground state decay rates were
taken as a lower bound to be used at low temperature and density.  At
higher temperature and density, the weak rates of
\citet{Ful80,Ful82a,Ful82b,Ful85}, \citet{Lan00}, and \citet{Oda94}
were used.  For details of the implementation see \citet{Wea78} and
\citet{Heg01}.

The outcome of presupernova evolution in this mass range is known to
be sensitive to the treatment of semi-convection and convective
overshoot mixing.  An important consequence is setting the CO-core
mass that results for a given main sequence mass (\Tab{endstate}).
Our CO-core masses turned out to be somewhat larger, for a given main
sequence mass, than in some other studies reflecting a greater
efficiency of overshoot mixing. For example, calculations using the
MESA code sometimes give smaller CO cores \citep{Tug14}.  There is
considerable variation for this quantity in the literature though
\citep{Eld04,Tug14}, and our core masses are within previously
published ranges.

Results can also be sensitive to zoning.  In most cases of interest
carbon, oxygen, or silicon burning ignite off center as convective
shells whose base is characterized by a large, nearly discontinuous
temperature inversion.  These burning shells propagate inwards in mass
by way of conductive flames or Rayleigh-Taylor instability.  Too large
a zone could artificially halt the migration of the flame
\citep{Tim94,Sie06} in a calculation that did not include a subgrid
model for the burning.  Typical calculations here used 1000 to 1300
Lagrangian zones, but the zoning was not uniform. Due to the rapid
temperature change near the flame, fine zones were automatically
inserted in its vicinity.  To avoid the runaway insertion of zones in
the temperature discontinuity, a minimum mass was specified.  Except
for a single sensitivity study at 9.5 \Msun, no zones smaller than $2
\times 10^{-3}\,\Msun$ were allowed, except near the surface of the
star.  In practice, this meant that zones in the vicinity of the flame
typically had a thickness of roughly $2$ to $3\,$km, except very near
the center where they were thicker.

\citet{Jon13} have suggested the possible importance of thermohaline
mixing in the study of stars in this mass range. Thermohaline mixing
is included in KEPLER using the formalism of \cite{Bra97} and
\citet{Kip80}, but adopted to a general equation of state as done for
the stability considerations described in \citet{Heg05} and using an
efficiency coefficient, $\alpha_{Th}$, of unity.  This formulation has
also been used by \citet{Can10}. We did not find that thermohaline
mixing had an important effect in any of our oxygen-burning or
silicon-burning flames since the laminar flame propagation and
Rayleigh-Taylor mixing dominated.

\section{PRESUPERNOVA EVOLUTION BELOW $9.0\,\Msun$}
\lSect{presn}

While the critical masses depend upon the code physics, especially for
convection, the evolution of massive stars below $9.0\,\Msun$ can be
segregated into three broad categories based upon their final outcome
(\Tab{endstate}): \textit{1)} CO white dwarfs (below $7.0\,\Msun$);
\textit{2)} ONe white dwarfs ($7.0\,\Msun$ - $9.0\,\Msun$, depending
upon mass loss); and \textit{3)} electron-capture supernovae (upper
end of $7.0\,\Msun$ - $9.0\,\Msun$, depending upon mass loss).

There have been many studies of this mass range.  \citet{Doh14} and
\citet{Jon13} recently carried out surveys similar to ours, though
neither followed the post-oxygen burning evolution.  Both found
outcomes qualitatively similar to ours, although certain critical
masses were shifted upwards in mass by about $1.0\,\Msun$ in the study
of \citet{Doh14}. They obtained CO white dwarfs for stars below
$8.0\,\Msun$, whereas our limit is $7.0\,\Msun$. For $8.0\,\Msun$
itself, they found a transition object that only partly burned its
carbon, quite similar to our $7.0\,\Msun$ model.  From $8.5\,\Msun$ to
$9.7\,\Msun$, they found ONe white dwarfs, and, at $9.8\,\Msun$, they
found electron capture supernovae.  We obtain ONe dwarfs and
electron-capture supernovae from $7.5\,\Msun$ to $8.75\,\Msun$, the
same upper bound as also found by \citet{Jon13}. The variation in
critical masses in these three studies can be attributed to the their
treatments of mass loss, semiconvection, convective overshoot mixing,
and, to a lesser extent, nuclear physics, opacities, and the initial
metallicity of the stars.  We claim no superiority for either
treatment, but point out that similar categories of behavior are
probable in any evolution code that carries all the relevant physics,
but with mass shifts due to uncertainties in that physics.

\subsection{Carbon-Oxygen White Dwarfs - M $< 7\, \Msun$}
\lSect{CO}

A representative case of a star that made a CO white dwarf was the
$6.5\,\Msun$ model.  This star had a main sequence lifetime of
$45\,$Myr and a helium burning lifetime of $12\,$Myr, identical to the
$6.5\,\Msun$, $Z = 0.02$ model of \citet{Doh14}.  Our star developed a
maximum helium convective core near helium depletion of $0.83\,\Msun$,
however, which resembles more their $7.5\,\Msun$ model (they found
$0.79\,\Msun$ for the maximum convective core in that model).  At
central helium depletion, our helium core had a mass of
$1.478\,\Msun$.  Later, convective dredge up reduced the helium core
and CO-core to $0.96\,\Msun$.  At that time, the helium shell became
very thin and flashes began.  Our star then had a luminosity of $3.66
\times 10^{37}\,$erg$\,$s$^{-1}$, a radius of $1.52\times
10^{13}\,$cm, and a total mass of $6.397\,\Msun$.  Its central
temperature and density were $9.1\times10^7\,$K and $2.5\times
10^7\,$g$\,$cm$^{-3}$, respectively, and its composition is given in
\Fig{comp6.5}.

% fig 1 - composition  6.5
\begin{figure}
\includegraphics[width=\columnwidth]{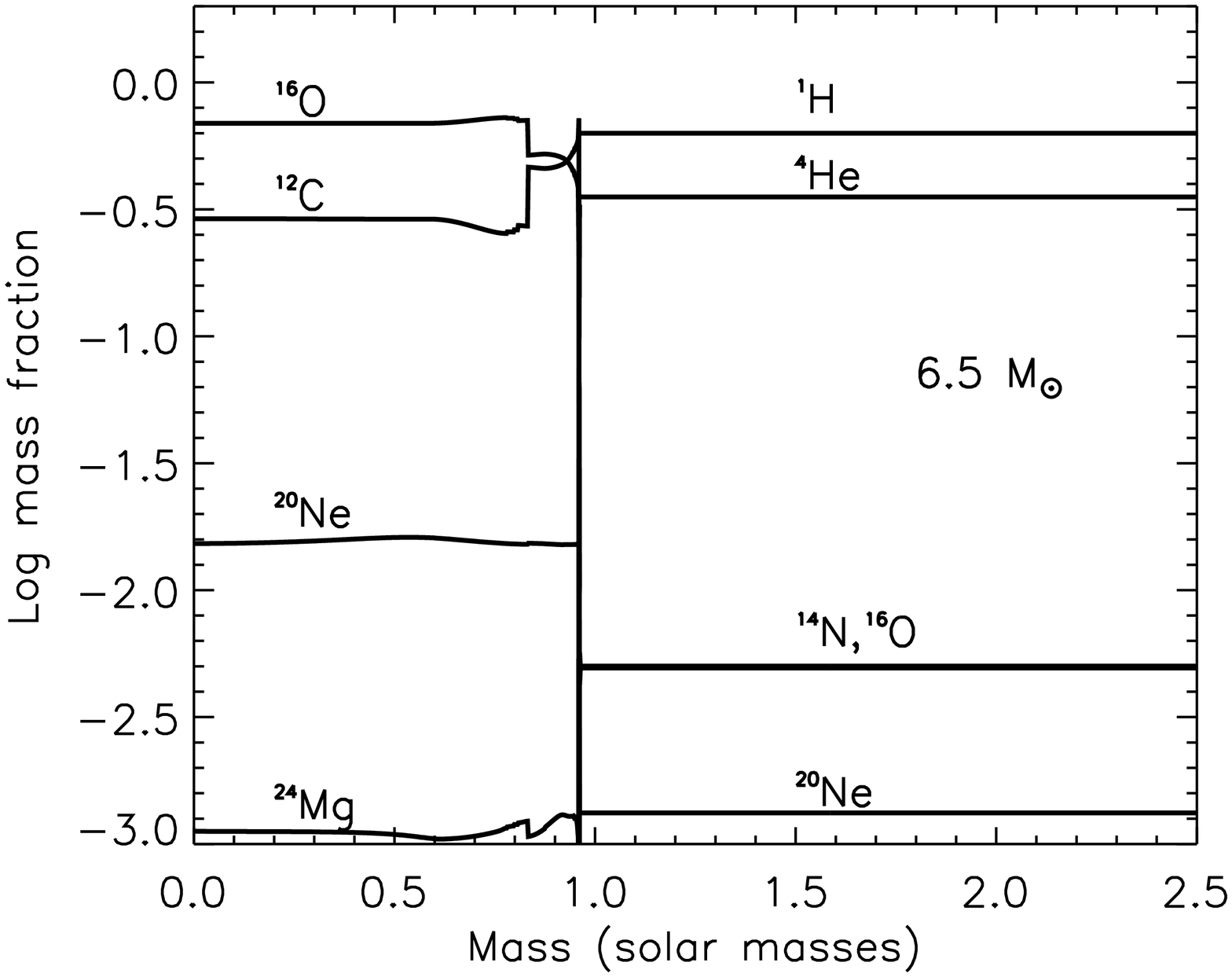}
\caption{Composition by mass fraction of the inner $2.5\,\Msun$ of the
  $6.5\,\Msun$ model at a time when nuclear burning has ceased in the
  core.  The central density is $2.45 \times 10^7\,$g$\,$cm$^{-3}$ and
  temperature, $9.1 \times 10^7\,$K.  The luminosity and radius are
  $3.94 \times 10^{37}\,$erg$\,$s$^{-1}$ and $1.59 \times 10^{13}\,$cm
  and the present star mass is $6.38\,\Msun$.  Thin shell flashes were
  not resolved in this study, but continued evolution will probably
  produce a carbon-oxygen white dwarf. \lFig{comp6.5}}
\end{figure}

The thin helium shell flashes were not resolved and the subsequent
evolution of this star was not followed.  For commonly employed mass
loss rates, the envelope will be lost before the CO core approaches
the Chandrasekhar mass and a white dwarf will result.  There is some
possibility that a few of these stars might become ``Type 1.5
supernova'' \citep{Ibe83} if the Chandrasekhar mass is reached before
the entire envelope is ejected.  Altogether, the $6.5\,\Msun$ model
here resembled closely the $7.0\,\Msun$ to $7.5\,\Msun$ models of
\citet{Doh14}, the chief difference with their $7.0\,\Msun$ model
being the greater extent of the helium convective region during core
helium burning.

The $7.0\,\Msun$ model itself was an interesting transition case which
ignited, but did not complete carbon burning.  Its hydrogen and
burning lifetimes were $38\,$Myr and $9.6\,$My and the maximum extent
of the helium convective core, $0.91\,\Msun$.  At central helium
depletion, the helium core mass was $1.60\,\Msun$.  Once again, these
lifetimes are the same as the $7.0\,\Msun$ model of \citet{Doh14}, but
the extent of the convective core was more like their heavier
$8.0\,\Msun$ model.  Carbon burning ignited off center at
$0.33\,\Msun$ when convective dredge up had reduced the helium plus CO
core mass to $1.04\,\Msun$.  This is slightly less than the minimum
core mass for carbon ignition of $1.06\,\Msun$ cited by \citet{Doh14}.
The density and temperature at the carbon ignition point were
$2.2\times10^6\,$g$\,$cm$^{-3}$ and $6.5\times10^8\,$K, similar to
values previously found by \citet{Sie06}.  $Y_\mathrm{e}$ was $0.4988$
at the ignition point reflecting the initial metallicity of the star
with no appreciable electron capture prior to that point.  A
convectively bounded carbon-burning flame moved into the center,
eventually giving the composition in \Fig{comp7.0}. The central
temperature and density for the last model calculated were
$8.0\times10^7\,$K and $4.2\,\times 10^7\,$g cm$^{-3}$. $Y_e$ was
still 0.4986.  Thin helium shell flashes were not resolved and the
subsequent evolution of the star was again not followed.  Overall the
evolution of this $7.0\,\Msun$ model was similar to the $8.0\,\Msun$
model of \citet{Doh14} and the $9.0\,\Msun$ model of \citet{Sie06}
(though one would have to go to the $10.0\,\Msun$ model of
\citet{Sie06} to find a star that ignited carbon so close to the
center).

% fig 2 - composition  7.0
\begin{figure}
\includegraphics[width=\columnwidth]{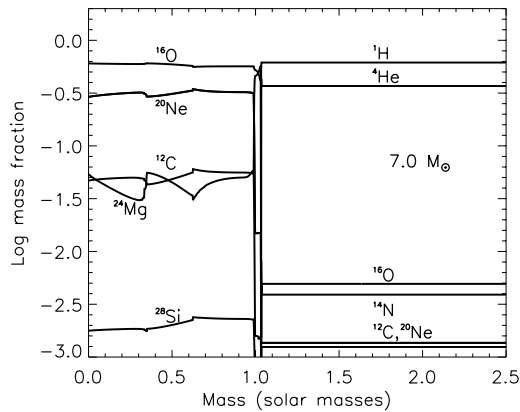}
\caption{Composition by mass fraction of the inner $2.5\,\Msun$ of the
  $7.0\,\Msun$ model at a time when nuclear burning has ceased in the
  core.  The central density is $4.23 \times 10^7\,$g$\,$cm$^{-3}$ and
  temperature, $8.1 \times 10^7\,$K.  The luminosity and radius are
  $4.49 \times 10^{37}\,$erg$\,$s$^{-1}$ and $1.71 \times 10^{13}\,$
  cm, and the present star mass at this time is $6.79\,\Msun$. This
  star is remarkable for having retained substantial unburned carbon
  in a core composed mostly of oxygen and neon when nuclear activity
  has ceased. If continued helium shell burning or accretion were
  ultimately to increase the degenerate core to the Chandrasekhar
  mass, it could possibly explode as a carbon-deflagration supernova.
  \lFig{comp7.0}}
\end{figure}

The degenerate CO core of the $7.0\, \Msun$ model still had 
$3\,\%$ - $10\,\%$ carbon, by mass fraction, remaining in its interior
(\Fig{comp7.0}).  If the envelope is not lost prior to reaching the
Chandrasekhar mass, this would be sufficient to ignite a carbon
deflagration and produce a Type 1.5 (single star) or Type~Ia
supernova (mass exchanging binary).

In summary, we find for this mass range, using the standard physics in
KEPLER, a value for the upper mass that does not ignite carbon burning
of approximately $7\,\Msun$.  This is about $1\,\Msun$ less than the
often cited value of $8\,\Msun$, but well within the range of values
obtained in other studies. For example, \citet{Bre93} find $M_{\rm
  up}$ of $5\,\Msun$ - $6\,\Msun$ and \citet{Poe08} find a value of
$8.5\,\Msun$ to $9.0\,\Msun$.  For other references and a discussion
see \citet{Sie06}.  The range is probably largely due to the varying
treatment of convective overshoot mixing and semiconvection by
different groups.

\subsection{Oxygen-Neon White Dwarfs and Electron-Capture Supernovae 
7.0 - $9.0\, \Msun$}
\lSect{NeO}

Moving on up in mass, one encounters stars that are able to deplete
carbon in their cores before becoming cold degenerate objects, but are
unable to ignite neon and oxygen burning.  Typical of stars in
this mass range is the $7.5\,\Msun$ model.  This star had a maximum
convective extent during helium core burning of $1.01\,\Msun$ and a
helium core mass at helium depletion of $1.75\,\Msun$, comparable to
the $8.5\,\Msun$ to $9.0\,\Msun$ model of \citet{Doh14}.  Our 7.5
\Msun \ model ignited carbon off center at $0.092\,\Msun$ where the
density and temperature were $2.2\times10^6\,$g$\,$cm$^{-3}$ and
$6.6\times 10^8\,$K, respectively.  Carbon burning moved inwards as a
CBF and, after reaching the center, was followed by episodes of carbon
shell burning, finally leaving a degenerate core with the composition
shown in \Fig{comp7.5} when all nuclear burning had ended.  The size
of the helium plus ONe core at that point was $1.11\,\Msun$.  The
evolution of the $8.0\,\Msun$ model was similar, but it and all
heavier models ignited carbon burning in the center.

% fig 3 - composition  7.5
\begin{figure}
\includegraphics[width=\columnwidth]{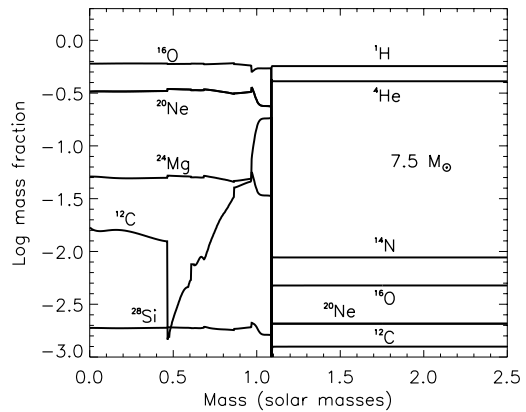}
\caption{Composition by mass fraction of the inner $2.5\,\Msun$ of the
  $7.5\,\Msun$ model.  The central density at this time is $6.33
  \times 10^7\,$g$\,$cm$^{-3}$ and the temperature,
  $6.8\times10^7\,$K.  The luminosity and radius are
  $5.11\times10^{37}\,$erg$\,$s$^{-1}$ and $1.92 \times 10^{13}\,$ cm,
  and the present star mass is $6.96\,\Msun$.  Too little carbon
  remains to greatly affect the subsequent evolution.  The end product
  will be an oxygen-neon white dwarf or, possibly, an electron-capture
  supernova. \lFig{comp7.5}}
\end{figure}

Further evolution would again have required the tracking of helium
shell flashes in very low mass zones and an uncertain mass loss
history, and was not attempted here. Other more thorough studies
\citep[e.g.][]{Poe08,Jon13,Doh14} suggest that stars in this mass
range will produce, in the common case, ONe white dwarfs.  For some
range of masses, depending upon the treatment of mass loss, the core
will grow to the Chandrasekhar mass before the envelope is lost and an
electron-capture supernova will result \citep{Jon13}.  The density
structure at the end when a supernova occurred would be similar to
that shown for the $8.75\,\Msun$ model in \Fig{presndn}.  By employing
coarse zoning that suppressed thin shell flashes, that model was
evolved until its core mass was $1.345\,\Msun$ and its central density
was $1.4\times10^9\,$g$\,$cm$^{-3}$, only a few hundredths of a solar
mass short of collapse.

% fig 4 -  density structure presn
\begin{figure}
\includegraphics[width=\columnwidth]{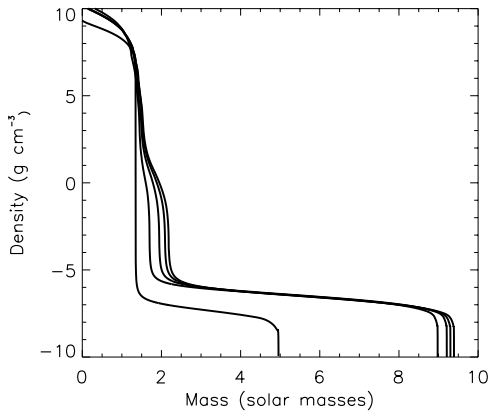}
\caption{The density structure of $5$ presupernova stars with main
  sequence masses of $8.75\,\Msun$, $9.25\,\Msun$, $9.5\,\Msun$,
  $9.6\,\Msun$, and $9.7\,\Msun$ (left to right).  All except the
  $8.75\,\Msun$ model are evaluated a the time of iron core collapse
  ($v_{\rm coll} =1000 $km$\,$s$^{-1}$).  These stars lost little mass
  prior to exploding and have very steep density declines outside the
  Chandrasekhar mass.  They ignited silicon burning without
  appreciable hydrodynamical mass ejection.  The $8.75\,\Msun$ model
  is evaluated when the central density has reached $2 \times
  10^9\,$g$\,$cm$^{-3}$ and the core mass is $1.345\,\Msun$.  It is
  composed chiefly of $^{16}$O ($52\,\%$) and $^{20}$Ne ($29\,\%$).
  It has already experienced appreciable mass loss and may end up a
  ONe white dwarf before collapsing as an ONe electron-capture
  supernova \citep{Nom84,Has93}. \lFig{presndn}}
\end{figure}

\section{CONVECTIVELY BOUNDED FLAMES (CBFs) ABOVE 9.0 \Msun}
\lSect{flame}

A major complication inhibiting the realistic modeling of heavier
stars in the 9.0 to $10.3\, \Msun$ mass range (\Tab{endstate}) is the
need to realistically represent the CBFs that ignite during their
post-carbon-burning evolution
\citep{Tim94,Ibe97,Sie06,Jon13,Jon14}. CBFs occur when any major fuel
(hydrogen, helium, carbon, neon, oxygen, or silicon) ignites off
center. Above 9.0 \Msun, hydrogen, helium and carbon ignite centrally
and it is the combustion of the heavier fuels - oxygen, neon, and
silicon - that complicates matters.  The temperature inversion that
causes off-center ignition is due to the interaction of degeneracy and
neutrino losses and, because of the high temperature sensitivity of
the burning processes, the energy transport above the burning region
is always convective.

The fuel first ignites somewhere off-center in a degenerate
flash. Depending upon the degree of the degeneracy, the flash can be
weak or violent, and the burning quickly drives convection out to the
mass necessary to radiate the excess energy as neutrinos. Initially,
the temperature at the base of the shell is not high enough to lead to
substantial inward motion of the burning front by conduction, though
there could, of course, be convective ``undershoot mixing''.  That is,
the low temperature in the temperature gradient immediately beneath
the convective zone initially implies such a slow conductive flame
speed that the time for the base to move inwards in mass is long
compared to the time for the composition in the shell to evolve. Thus,
initially, the base of the shell stays approximately fixed in
Lagrangian coordinate while the fuel burns in a shell. As the
temperature at the base of the convective shell rises in response to
the decreasing mass fraction of fuel though, the conductive flame
speed, which is very sensitive to that ``bounding temperature'',
accelerates inwards. In the resulting flame, burning in the
temperature gradient just beneath the convective shell raises the
local temperature to a critical value where nuclear energy generation
exceeds radiative and conductive losses. That region then flashes to a
higher temperature that is super-adiabatic compared with the next zone
out. Unspent fuel spills into the convection shell and is mixed
outwards, providing a new source of fuel for burning in the shell. The
energy produced is mostly lost to neutrinos. The next zone in heats up
by conduction and the process repeats. As the flame moves into the
unburned matter, the decreasing area bounded by the flame requires the
flame to move faster, hence it becomes hotter.

\subsection{A Microscopic View of CBFs}
\lSect{microflame}

There are two ways of viewing such flames. The ``microscopic view'',
which has just been described, envisions the flame as a thin sheet
with a specified boundary temperature that propagates locally, by
conduction, into a homogeneous medium with prescribed properties like
a modified laminar flame \citep[e.g.][]{Tim94}. The other
``macroscopic'' view (\Sect{macroflame}) views the flame as a boundary
condition the star adjusts to accommodate the global constraints of
hydrostatic equilibrium, composition, and neutrino losses. The two
views are complementary ways of describing the same physical
phenomenon.

In the microscopic view, if the flame were well resolved (it isn't
usually), the process of burning and mixing would be continuous.  The
energy generation resulting from the flame's propagation would vary
slowly according to the local composition, density, and bounding
temperature. The flame properties would agree with those estimated by
evaluating the usual laminar flame speed formula \citep{Lan59} with
appropriate modifications,
\begin{eqnarray}
v_{\rm flame} &\approx \left(\frac{c \epsilon}{\kappa \rho
  E} \right)^{1/2} \\
\delta_{\rm flame} &\approx \left(\frac{c E}{\kappa \rho
  \epsilon}\right)^{1/2} 
\lEq{flamespeed}
\end{eqnarray}
where $v_{\rm flame}$ is the flame speed, $\delta_{\rm flame}$, its
width, $\epsilon$, the nuclear energy generation rate, $\kappa \rho$,
the reciprocal mean free path, c, the speed of light, and $E$, the
energy difference, in erg g$^{-1}$, between the cold fuel beneath the
flame and the base of the convective shell. Results are most sensitive
to the energy generation, and it is here that things can become
complicated. For a ``free flame'' the temperature and energy
generation can rise without bound and the above equations give a very
narrow, fast flame \citep{Tim92}. At 10$^{9}$ g cm$^{-3}$, for
example, the flame speed and width for an oxygen-neon burning flame
are 6.4 km s$^{-1}$ and $1.5 \times 10^{-3}$ cm. For the CBF, however,
the burning temperature cannot exceed the bounding temperature. In
fact, the temperature where conduction inwards balances energy
generation is a bit less than the bounding temperature \citep[][their
  Fig. 8]{Tim94}. Thus a CBF moves much more slowly and is broader
than an unbounded or ``free'' flame.

\subsubsection{Oxygen Burning CBFs}
\lSect{oflame}

A case of particular interest here is a CBF moving into a core of
about 60\% oxygen and 30\% neon. Eventually, both oxygen and neon
burn, but in the conductive region that sets the flame speed, only a
little fuel has burned, and that is chiefly neon.  Since neon burning
proceeds by a photodisintegration-rearrangement reaction, the energy
generation is, to first order, independent of the density.  Taking a
temperature of $1.8 \times 10^9$ K gives an energy generation rate in
the neon-oxygen material in KEPLER using the 220 isotope network of
$\sim10^{14}$ erg g s$^{-1}$.  Order of magnitude estimates for the
other quantities in \Eq{flamespeed} at an appropriate time in, e.g., a
9.5 \Msun \ model are $\kappa \rho \sim 5 \times 10^5$ cm$^{-1}$ and
and $E \sim 2 \times 10^{16}$ erg g$^{-1}$. Both quantities depend
only weakly upon the temperature and density.  Using these values in
\Eq{flamespeed} gives a crude estimate of the flame speed of $20$ cm
s$^{-1}$ and its width, 3500 cm.

More accurate results have been obtained off-line from the highly
resolved numerical modeling of parametrized flames. Such studies have
been done for oxygen-neon flames \citep{Tim94}, but not for silicon
burning. For the oxygen-neon flame described above, these studies
find, for $T_{\rm bound} = 2 \times 10^9$ K, a flame speed $v_{\rm
  flame}$ = 1.91 cm s$^{-1}$ and width, $\delta_{\rm flame}$ = 250 cm,
very much smaller than any practical stellar zone
thickness. Apparently evaluating the energy generation at 90\% of the
bounding temperature though, 1.8 vs 2.0 $\times 10^9$ K, gives the correct
order of magnitude. Changing the bounding temperature used in the
estimate would greatly distort the speed-to-width ratio, though one
could artificially change the opacity and achieve agreement. The chief
use of \Eq{flamespeed} is in understanding and extrapolating the more
accurate numerical studies. For example, in Table 6 of \citet{Tim94}
the flame speed depends on the bounding temperature roughly as
T$^{18}$. This is the approximately the square root of the temperature
dependence of the energy generation rate in KEPLER in the relevant
temperature range. Similarly the width goes as T$^{-18}$ and both are
approximately independent of the density.

In order to follow oxygen-neon flames in KEPLER, a subgrid
representation of the propagation was implemented.  First the location
of the flame on the grid, if there was one, was determined based upon
composition and temperature information. A flame was deemed to exist
when, moving outwards, a stellar zone was encountered where the
temperature increased by more than 10\% and the mean atomic weight
decreased by more than 2\%. In addition, the hotter zone was required
to be hotter than $1.5 \times 10^9$. For cooler temperatures the flame
speed is negligible. This algorithm, which was only used for neon and
oxygen burning, proved successful in unambiguously determining the
location of the burning front. Energy was then artificially deposited
in the cooler underlying zone on a time scale given by the thickness
of the zone, $\Delta r$, divided by the desired flame speed, $v_{\rm
  o-cbf}$. The effective energy generation rate in the cool zone, j,
was then
\begin{equation}
\epsilon_f \ = F \frac{(q_{j+1} - q_j) v_{\rm o-cbf}}{\Delta r}
\lEq{oflame}
\end{equation}
where $q_j$ is the internal energy in erg g$^{-1}$ of the cooler zone,
$q_{j+1}$, the internal energy of the hotter zone, F, a multiplier of
order unity, and $\epsilon_f$, the artificial energy generation rate in
erg g$^{-1}$ s$^{-1}$. This was added to any nuclear energy generation
in zone j. In order to conserve energy and maintain code stability an
equal amount of energy was subtracted each time step from the hot
zones, j+1 though j+5. In practice, convection kept those zones
coupled to the larger heat reservoir of the convective shell
surrounding the flame. In response to the artificial energy
deposition, the cooler zone was gradually heated to some flash point
where its nuclear energy generation exceeded $\epsilon_f$ in
\Eq{oflame}, after which the temperature in the zone rapidly ran away
until it became convectively linked with zone j+1. The flashing of
individual zones resulted in a sputtering flame, but one whose average
speed was determined by $v_{\rm o-cbf}$ and F. The parameter F was
adjusted to make the flame move at a value close to the actual value
of $v_{\rm o-cbf}$, which empirically required $F \approx 2$.

It remained only to prescribe $v_{\rm o-cbf}$. Fortunately, in the case
of oxygen burning CBFs, there exist off-line studies that gave the
flame speed as a function of bounding temperature (the temperature of
zone j+1) quite accurately. These are given in Table 5 of
\citet{Tim94}. While this table is for a single composition, the
actual composition in our stars was not far from 60\% oxygen, 40\%
neon and the results do not depend sensitively upon the actual
mixture. The density range of the table, 2 - $10 \times 10^8$ g
cm$^{-3}$, needed to be extended down slightly to a few $\times 10^7$
g cm$^{-1}$, but fortunately, as discussed above, the oxygen-neon CBF
is not sensitive to density. A fit to Table 5 of Timmes et al. gave
\begin{equation}
v_{\rm o-cbf} \ \approx \ 40 \left(\frac{T_9}{2.5}\right)^n \ {\rm cm \ s^{-1}}
\lEq{Timmes}
\end{equation}
where n is 18 in the range $1.5 \leq T_9 < 2$, 13, for $2 \leq T_9 <
2.5$, and 10 for $T_9 \> 2.5$. In practice, the subgrid model gave a
flame that moved at a speed that was typically within a factor of two
of this value. 

While this approach worked well in a 1D stellar evolution code, the
actual stellar physics is certainly more complicated. The flame probably
remains thin, but the fuel-ash interface may significantly diverge
from spherical symmetry. In particular, Rayleigh-Taylor instability or
convective undershoot mixing might significantly accelerate the
burning. The laminar CBF speed used here should thus be regarded as a
lower bound to the actual burning rate in 3D.

\subsubsection{Silicon burning CBFs}
\lSect{siflame}

Unfortunately there are no corresponding tables for the propagation of
a CBF in dense, neutronized silicon and sulfur.  A rough estimate of
the flame's properties can be obtained using \eq{flamespeed}. While
the bounding temperature changes significantly with time, a
representative condition might be $4.5 \times 10^9$ K for a density $5
\times 10^8$ g cm$^{-3}$. For a neutron-rich mixture of silicon and
sulfur isotopes, mostly $^{30}$Si and $^{32}$Si. the nuclear energy
generation rate from KEPLER at a temperature near 90\% of $T_{\rm
  bound}$ is $\sim5 \times 10^{17}$ erg g s$^{-1}$. Taking $\kappa
\rho \sim 10^6$ cm$^{-1}$, and $E \sim 2 \times 10^{16}$ erg g$^{-1}$
gives a flame width of about 30 cm and speed of 9 m s$^{-1}$,
considerably faster than the oxygen-neon CBF described above.  A fine
zoned calculation of a small sphere (59\% $^{30}$Si, 34\% $^{32}$Si,
3.5\% $^{34}$S, and 3.5\% $^{36}$S; $Y_e$ = 0.456) at constant
pressure with zoning of about 1 cm using the technique of
\citet{Nie97} gave a width of about 10 cm and a speed of 2 m
s$^{-1}$. Both estimates are order of magnitude guesses and the real
speed will vary considerably with the bounding temperature, but the
estimates show that the silicon flame is also very thin and moves
rapidly.

Fortunately, the speed of the silicon burning CBF is not crucial to
the most important conclusion of this paper, namely the dynamical
nature of the silicon flash in stars around 10 \Msun. Also, other
physics comes into play that may render the propagation speed of a
laminar silicon flame moot. The burning interface is increasingly
Rayleigh-Taylor unstable because of the inversion of mean atomic
weight and $Y_e$, and this instability also causes the rapid
advancement of the burning. 

Following the silicon flash in stars from 9.0 to 10.2 \Msun, there
remains a core of unburned silicon of up to 0.4 \Msun \ (much less in
the heavier models; \Tab{siflash}). This silicon nodule is surrounded
by a thick shell of iron that has experienced substantial electron
capture. The silicon itself is also neutron rich, having been made by
an oxygen CBF operating at high density, but not so neutronized as the
iron above it. Representative values for the electron mole number are
$Y_e$ = 0.465 for the silicon and 0.45 or less for the iron. The mean
atomic mass is also inverted with $\bar A \approx 30$ in the silicon
and 54 in the iron. Were it not for the accompanying temperature
inversion, the iron-silicon interface would clearly be Rayleigh-Taylor
unstable.

The temperature is inverted, though, with $T_9 \approx 4$ to 5 in the
iron ash, once the core has relaxed and reignited silicon burning, and
$T_9 \approx 2.5$ in the silicon. This temperature inversion prohibits
mixing so long as the bounding temperature stays above a critical
value. Neutrino losses are continually seeking to cool the convective
shell and erode this barrier. The iron is kept hot by the small influx
of silicon that is continually being mixing outwards from the silicon
CBF.  Were this mixing to stop, the silicon abundance would decrease in
the convective shell, the temperature would go down, just a bit, and
the interface would become Rayleigh-Taylor unstable. This would mix
silicon out and rekindle the burning.

Employing the Ledoux criterion for convection, the solution the KEPLER
code finds to this constraint is to, rarely and briefly, convectively
link a single cooler silicon zone with an overlying hotter convective
zone of iron. A small amount of mixing powers a brief resurgence of
burning in the convective shell that raises its temperature slightly
above the threshold required for instability by the Ledoux criterion
and shuts off the convective link with the inner zone. Once the bit of
fuel that was mixed outwards gets depleted, the temperature in the
convective shell decreases and the convective linkage occurs
again. Backups in the code prevent a large amount of fuel from mixing
during a single episode.

Over time, the mixing also raises the temperature in the cooler zone
by bringing in hot ash. Eventually, its temperature increases to the
point where the cooler zone itself runs away and became permanently
convective. The burning in KEPLER is thus self-regulating. If too
little fuel is mixed into the burning front, the convection shuts off
and the bounding temperature declines, making the inversion in $\bar
A$ and $Y_e$ more effective and causing new mixing. If too much is
mixed, the bounding temperature rises and shuts off mixing.

While this behavior reflects a real mixing instability, its
implementation is debatable. The burning front advances because of
intermittent convection, not conduction. What does it mean for
convection to turn on and off on sub-millisecond time scales - far
faster than e.g., sound can cross a zone?  Basically, it is the star's
response to a real macroscopic dilemma - how to keep burning when the
only available fuel lies beneath a temperature inversion that acts to
stabilize it against mixing. However the mixing occurs, it must
continue at a rate sufficient to at least balance the star's neutrino
losses. Throughout the mixing of a single zone, the global neutrino
losses, $Q_{\rm \nu}$, stay relatively constant. The average flame
speed is then
\begin{equation}
v_{\rm flame} \ \approx \ Q_{\rm \nu}/(4 \pi r_{\rm shell}^2 \, \rho \,
q_{\rm nuc}),
\lEq{flameeq}
\end{equation}
where $q_{\rm nuc}$ is the energy yield per gram from burning a given
fuel - oxygen or silicon, $ r_{\rm shell}$ is the radius of the flame,
and $\rho$, the local density.  For typical conditions, $Q_{\rm \nu}
\sim$ 10$^{44}$ erg s$^{-1}$, $r_{\rm shell} \sim 10^8$ cm, $q_{\rm
  nuc} \sim 5 \times 10^{17}$ erg g$^{-1}$, and $\rho \sim 10^8$ g
cm$^{-3}$, the necessary flame speed is $v_{\rm flame} \sim 20$ cm
s$^{-1}$. This is not very different from that expected for a
conductive flame had one been resolved, based on the relations in
\Sect{siflame}, and perhaps both processes operate. An analytic subgrid
model for silicon CBFs was not included in the present study however.

\subsection{A Macroscopic View of CBFs}
\lSect{macroflame}

Consider a degenerate stellar core with a mass slightly in excess of
the cold, non-rotating Chandrasekhar mass for its given composition,
especially for its distribution of $Y_e$. Ignoring rotation, such a
configuration can only be stabilized against collapse by the presence
of finite entropy. An effective Chandrasekhar mass can be defined
\citep[e.g.,][]{Bar90} as
\begin{eqnarray}
M_{\rm eff} \ &\approx \ M_{\rm Ch}^0 \ \left(1 \ + \ \left(\frac{\pi
  k T}{\epsilon_F}\right)^2 \right)\\ 
& \approx \ M_{\rm Ch}^0 \ \left(1 \ + \ \left(\frac{s_e}{\pi
  Y_e}\right)^2 \right),
\lEq{chandra}
\end{eqnarray}
where $\epsilon_F$ is the Fermi energy, roughly $1.11 (\rho
Y_e)^{1/3}$ MeV for a relativistic degenerate gas, and $s_e$ is the
electronic entropy, roughly $\pi^2 k T/\epsilon_F$.
For a given core mass and $Y_e$ then, there is a minimum
entropy required to stabilize the core against collapse.

Now consider now a core where electron capture in a convective burning
shell has reduced $Y_e$ substantially below 0.50 (a value of 0.46 to
0.48 might be typical) in a large fraction of the mass. This happens
in oxygen and silicon burning, but not in carbon burning. Assume
further that before the capture ensued, the core already had a mass
very close to the cold Chandrasekhar mass for $Y_e = 0.50$, i.e., 1.39
\Msun, including corrections for relativity and Coulomb forces. In the
absence of nuclear burning, a core with the new $Y_e$ would collapse,
as also noted by \citet{Jon13}. Within this core however, there is a
convective burning shell with base temperature $T_{\rm bound}$ and
extent $M_{\rm conv}$. Because the temperature gradient in the
convective shell is adiabatic, the extent of the shell depends mainly
on $T_b$, so there are really just two variables, $T_b$ and the
location in mass of the burning shell. For a given CBF location, the
bounding temperature must be at least adequate to stabilize the star
against collapse.

At the necessary high temperatures, neutrino losses by the pair and
plasma processes (mostly the former) will attempt to drive the
temperature down and make the core unstable to collapse. The
instability is exacerbated by further electron capture within the
convective shell which reduces $Y_e$. Energy generation to balance
these neutrino losses can only be obtained by the advancement inwards
of the CBF. Much as in ``ordinary'' massive stars with stationary
oxygen burning shells, the burning seeks to maintain, on the average,
an overall condition of ``balanced power'' \citep{Woo02} with neutrino
losses balanced by nuclear energy generation. Here however, the
temperature-sensitive rate at which the flame moves plays the role of
the nuclear energy generation rate. The star seeks and finds a
solution where the flame moves at the necessary speed to balance
neutrino losses in the convective shell.  Not too surprisingly since
the mass fraction of the oxygen in the convective shell outside the
CBF is non-trivial, typically from 1\% to 10\% by mass, the bounding
temperature at which balanced power is achieved is similar to that for
ordinary oxygen shell burning, roughly $2.0 \times 10^9$ K. The flame
thus moves at a speed $\sim 1 $ cm s$^{-1}$ and the neutrino losses
are typically 10$^{42}$ - 10$^{43}$ erg s$^{-1}$.

The silicon burning flame stabilizes at a larger temperature and flame
speed where intermittent Ledoux convection mixes out enough silicon to
provide a global neutrino loss rate of about 10$^{45}$ erg
s$^{-1}$. This is what is required if silicon is to burn in a
convective shell with temperature at its base 4 to $5 \times 10^9$ The
flame neither accelerates rapidly nor extinguishes, but obeys, on the
average, a condition of balanced power.

\section{OFF-CENTER OXYGEN BURNING AND SILICON FLASHES  - 
9.0 TO $10.3\, \Msun$}
\lSect{sishell}

We are now equiped to discuss the actual models in the 9.0 to 10.3
\Msun \ range.  Above 9.0 \Msun, for the assumed physics, the
evolution of a massive star short of the pair-instability limit
culminates in the production of an iron core in hydrostatic
equilibrium that collapses to a neutron star or black hole. For stars
from 9.0 through $10.3\, \Msun$, the effects of degeneracy are still
very strong and, while carbon ignites centrally, neon and oxygen do
not. Silicon too usually ignites off center, (\Tab{siflash}), in a
powerful flash. At the upper end of the mass range (9.8 - 10.3 \Msun),
the flash is robustly violent enough to lead to a localized
thermonuclear runaway, a ``silicon deflagration''
(\Sect{siflash}). Deflagration may also occur for other masses in this
range (e.g., the 9.0 and 9.3 \Msun \ models), but the occurrence at
lower mass is more sensitive to the treatment of the flame. A 9.0
\Msun \ model which used an artificially slow oxygen burning flame did
not experience a deflagration. \Fig{rhoctc} summarizes the final state
of three stars, two of which experienced a very strong silicon flash
culminating in deflagration (9.0 and 10.0 \Msun \ models) and one of
which evolved ``normally'' igniting all burning stages at its canter
(11.0 \Msun \ model). \Fig{ye} shows substantial electron capture
occuring the in oxygen burning CBF in the 9.5 and 10.0 \Msun \ models.

% old fig 5 was 9.0 composition - no longer valid
% new fig 5 is t-rho-center for 9, 10, 11
\begin{figure*}
\begin{center}
\includegraphics[width=0.475\textwidth]{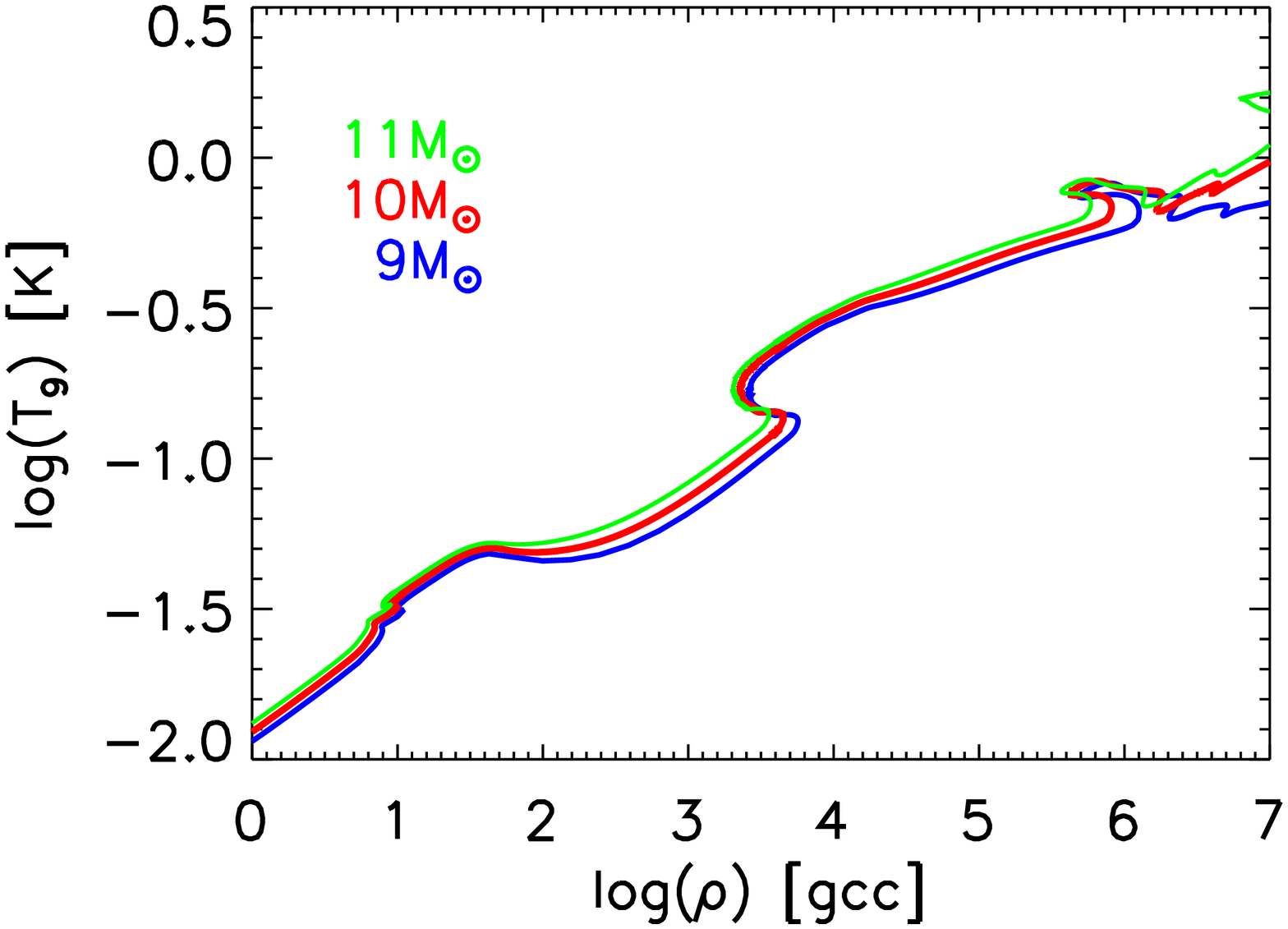}
\hfill
\includegraphics[width=0.475\textwidth]{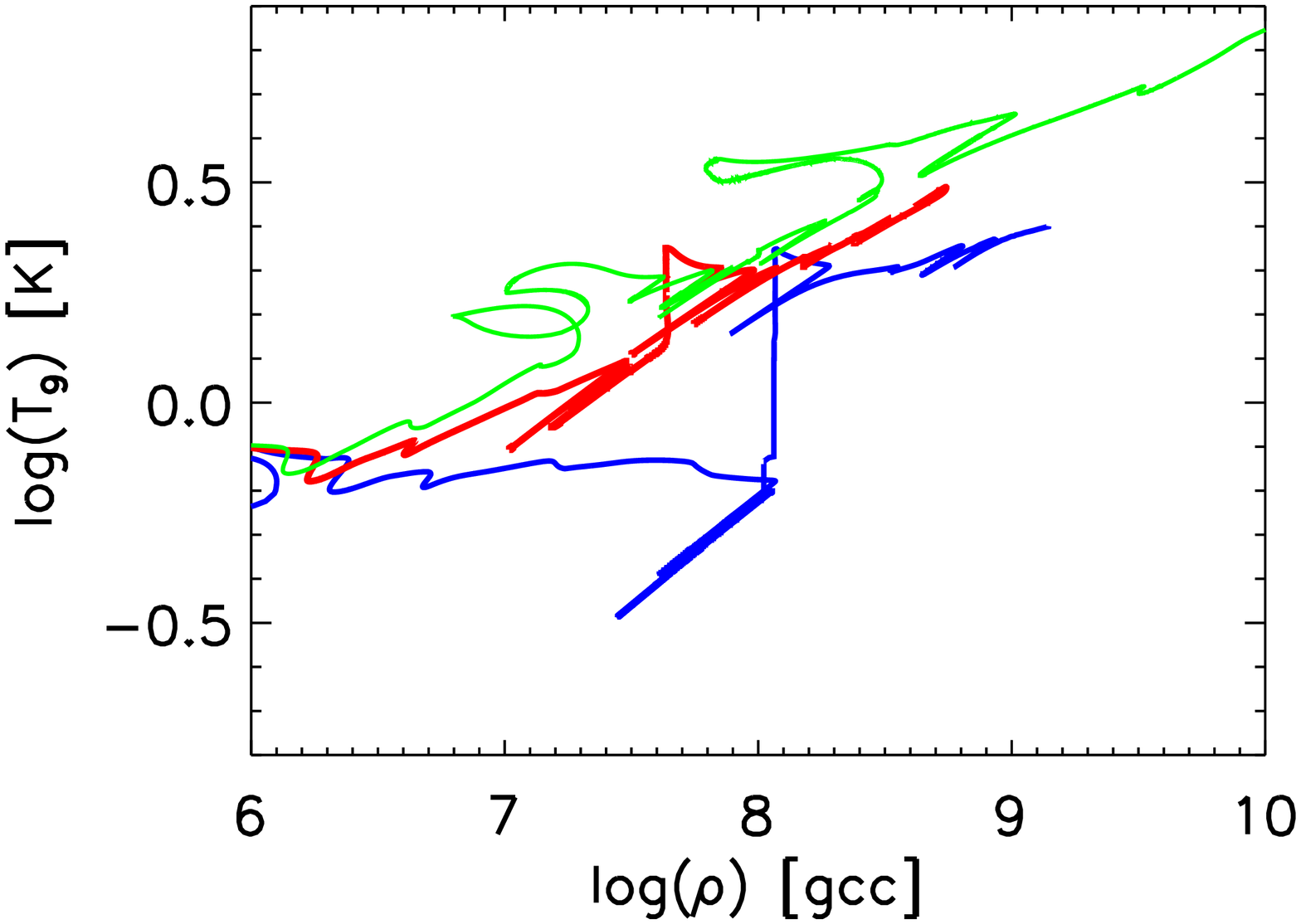}
\caption{Evolution of the central temperature and density for the 9.0,
  10.0 and 11.0 \Msun \ models. For the 9 and 10 \Msun \ models, the
  curves stop at the time of off-center silicon ignition. The top
  frame shows very similar evolution for all three stars through
  carbon burning ($\rho_c \approx 10^6$ and $T_9 = 0.8$). Beyond this
  point, the bottom frame shows divergence. The 11 \Msun \ model
  ignites and burns oxygen and silicon in its center and
  ends its life as a normal iron-core collapse supernova.  Owing to
  their greater degeneracy, the 9 and 10 \Msun \ models do not heat up
  as rapidly in the center after carbon depletion and ignite oxygen
  and neon burning off center. The diagonal sloping decline in density
  and temperature for the 9 and 10 \Msun \ models, starting at
  log $\rho_c$ near 8.0 and 7.6 respectively, result from the gradual
  expansion of the central regions in response to the off-center
  oxygen burning CBF. As the flame nears the center, less energy is
  being generated in the shell, however, and the core begins to
  contract and heat up again along the same path it followed when
  expanding. The abrupt upturn in temperature marks the arrival of the
  flame at the star's center. Other variations are caused by multiple
  shell burning episodes farther out in the star. While the central
  density is much higher at silicon ignition for the 9 \Msun \ model,
  the actual ignition occurs farther out in the star than for the 10
  \Msun \ model and the local ignition densities are actually very
  similar (\Tab{siflash}). \lFig{rhoctc}}
\end{center}
\end{figure*}

% fig 6 - ye evolution 9.5 and 10.0 Msun 
\begin{figure}
\begin{center}
\includegraphics[width=0.475\textwidth]{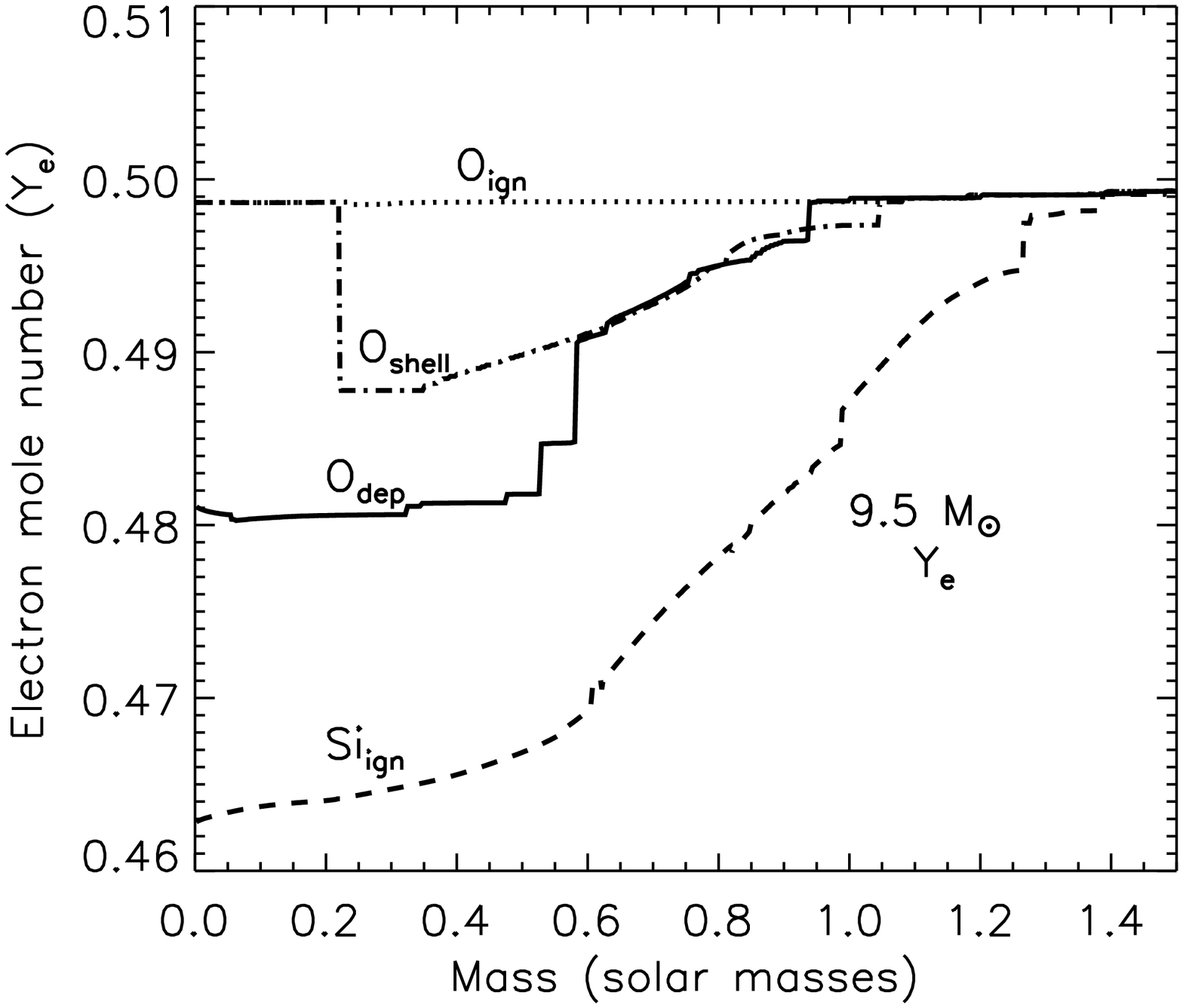}
\hfill
\includegraphics[width=0.475\textwidth]{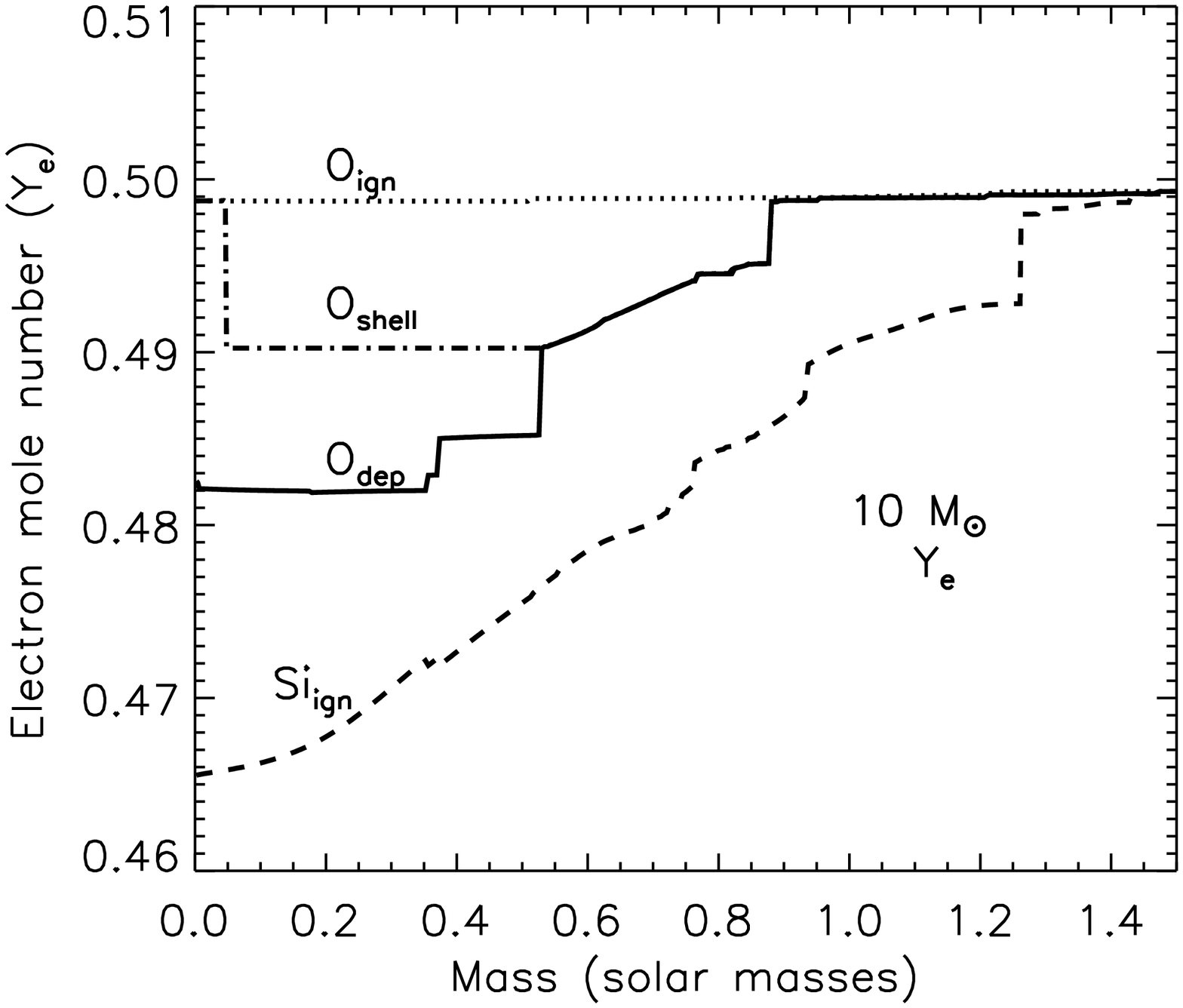}
\caption{Evolution of the electron mole number, $Y_\mathrm{e}$, in the
  inner $1.5\,\Msun$ of the 9.5 and 10.0 \Msun \ models.  Very little
  electron capture occurs in any of the cores prior to oxygen
  ignition, so $Y_\mathrm{e}$ is close to $0.50$ (dotted line).
  Electron capture occurs as the oxygen burning flame moves inwards,
  but oxygen depletes in the center of the star before silicon ignites
  (\Fig{comp9.5}). ``O$_{\rm shell}$'' corresponds to a time when the
  oxygen mass fraction first declines below about $10\,\%$ in the
  convective shell.  O$_{\rm dep}$ (the solid line) is when the oxygen
  mass fraction first goes below $5\,\%$ in the center of the star.
  Substantial neutronization occurs as the core contracts between
  oxygen depletion and silicon ignition. In both cases, a network of
  approximately $220$ isotopes was used to track the nuclear energy
  generation and weak interactions. \lFig{ye}}
\end{center}
\end{figure}

\subsection{Strong Silicon Flashes - No explosion}
\lSect{noexp}

Consider the illustrative case of the 9.5 \Msun \ model. Twenty-one
years prior to iron core collapse, neon ignites in a gentle ``flash''
0.252 \Msun \ (1560 km) off center. The neon burns in a growing
convective zone that eventually extends to 1.1 \Msun.
%s9.4#neign on /Volumes/DATA1/q/qdat/woosley/stars/tensol14/9.5/fine
Briefly, neon burning develops a power $\sim 10^{43}$ erg, but after
about a month, the power declines to 10$^{42}$ erg s$^{-1}$, where it
remains for about 10 years.  When the neon in this shell has been
depleted, the temperature at the shell's base rises to $1.65 \times
10^9$ K and a neon-burning CBF develops and begins to move
inwards. The flames velocity is very slow though, and three years later, it
has only reached 0.217 \Msun \ while its temperature has climbed to
$1.83 \times 10^9$ K, sufficient for oxygen burning. Off-center oxygen
burning thus commences with another mild flash 9.5 years before iron core
collapse (\Fig{comp9.5} and \Fig{flames}). The burning initially
powers convection out to 0.85 \Msun. A combined neon-oxygen burning
CBF then begins its advance into the core while the extent of the
convective oxygen burning shell shrinks in mass. About 5 years before
core collapse, oxygen has been only partly depleted in the convective
shell and with 3.6 years remaining, the temperature at the CBF becomes
high enough for the oxygen convective shell to briefly grow again. There
follows a resurgence of burning resulting from the rapid
regrowth of the convective shell out to 0.9 \Msun. After burning out
the oxygen in that shell, the CBF resumes its inwards progress
eventually arriving at the center 1.1 years before the iron core
collapses.

% fig 7 - composition 9.5
\begin{figure}
\begin{center}
\includegraphics[width=0.475\textwidth]{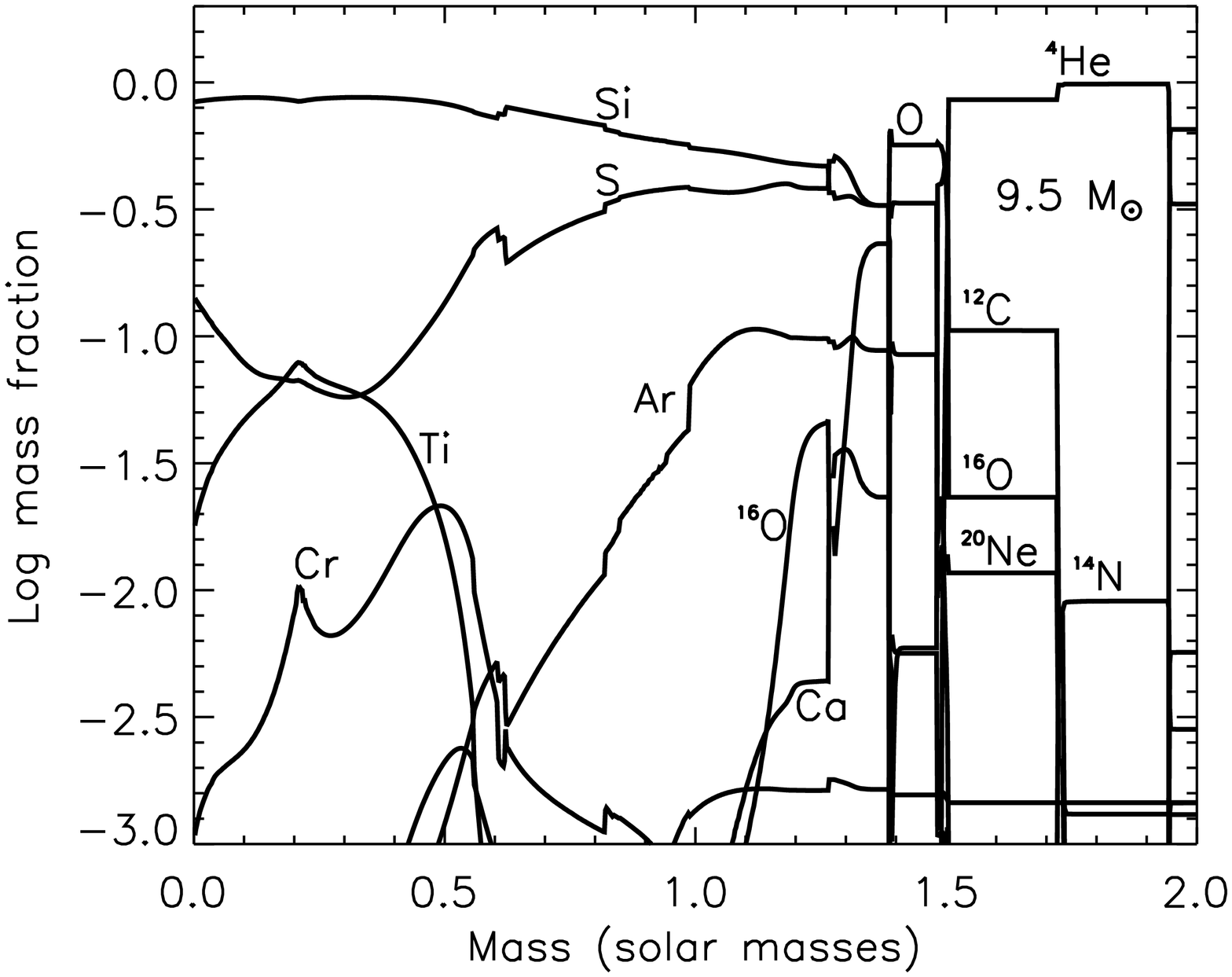}
\hfill
\includegraphics[width=0.475\textwidth]{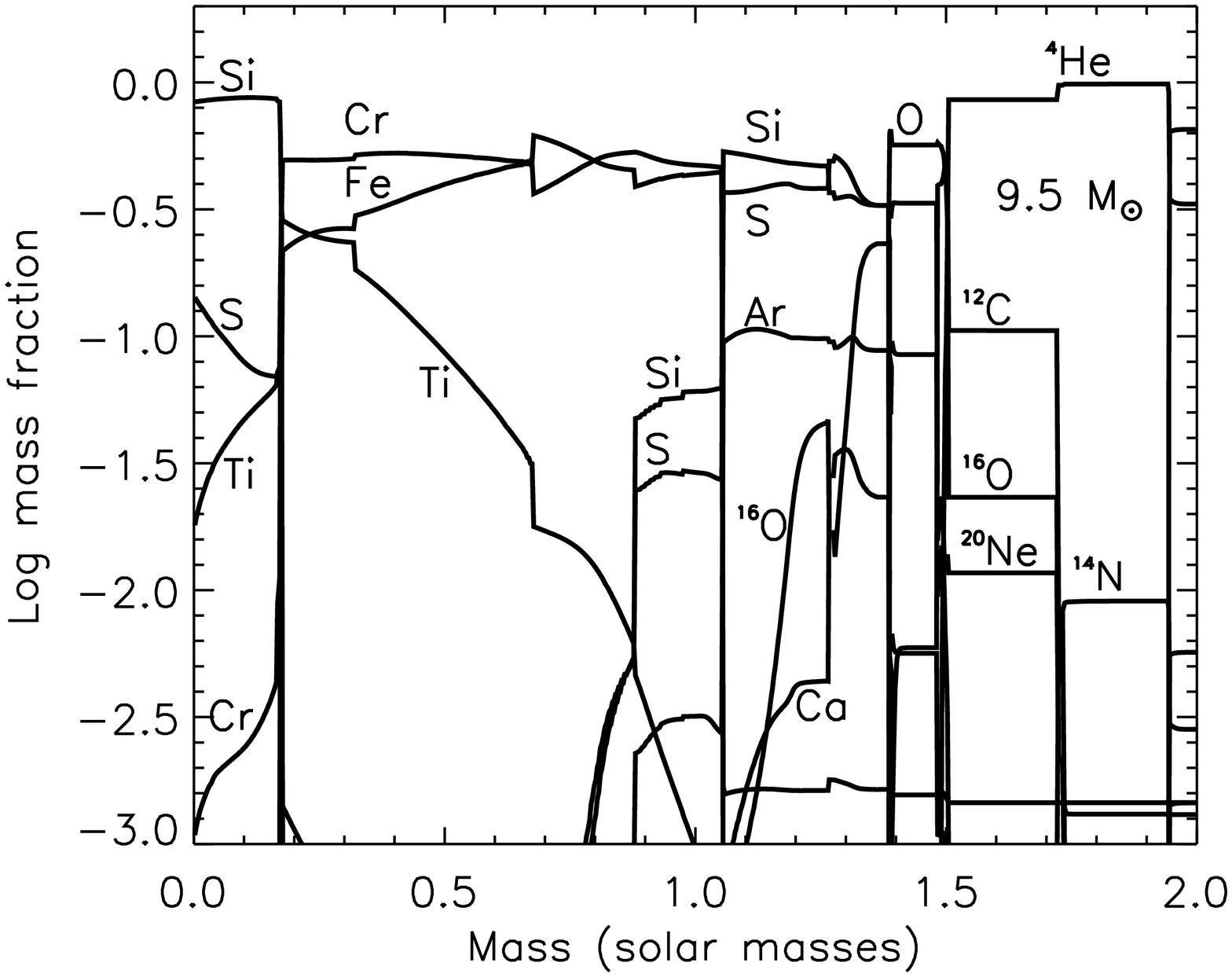}
\caption{Abundances in the $9.5\,\Msun$ model at silicon ignition
  (top) and during silicon shell burning (bottom).  Silicon burning
  ignites at $0.237\,\Msun$ at a density of $4.67 \times
  10^8\,$g$\,$cm$^{-3}$ and quickly establishes a convectively bounded
  flame.  The shell burning density and temperature in the bottom frame
  are $4.4 \times 10^8\,$g$\,$cm$^{-3}$ and $4.2 \times 10^9\,$K.
  Fifteen days after ignition, silicon burning reaches the center of
  the star and $1.8$ hours later the $1.29\,\Msun$ iron core
  collapses. \lFig{comp9.5}}
\end{center}
\end{figure}

% fig 8 - energy generation and convection in flames for 9.5 Msun
\begin{figure*}
\includegraphics[width=\textwidth]{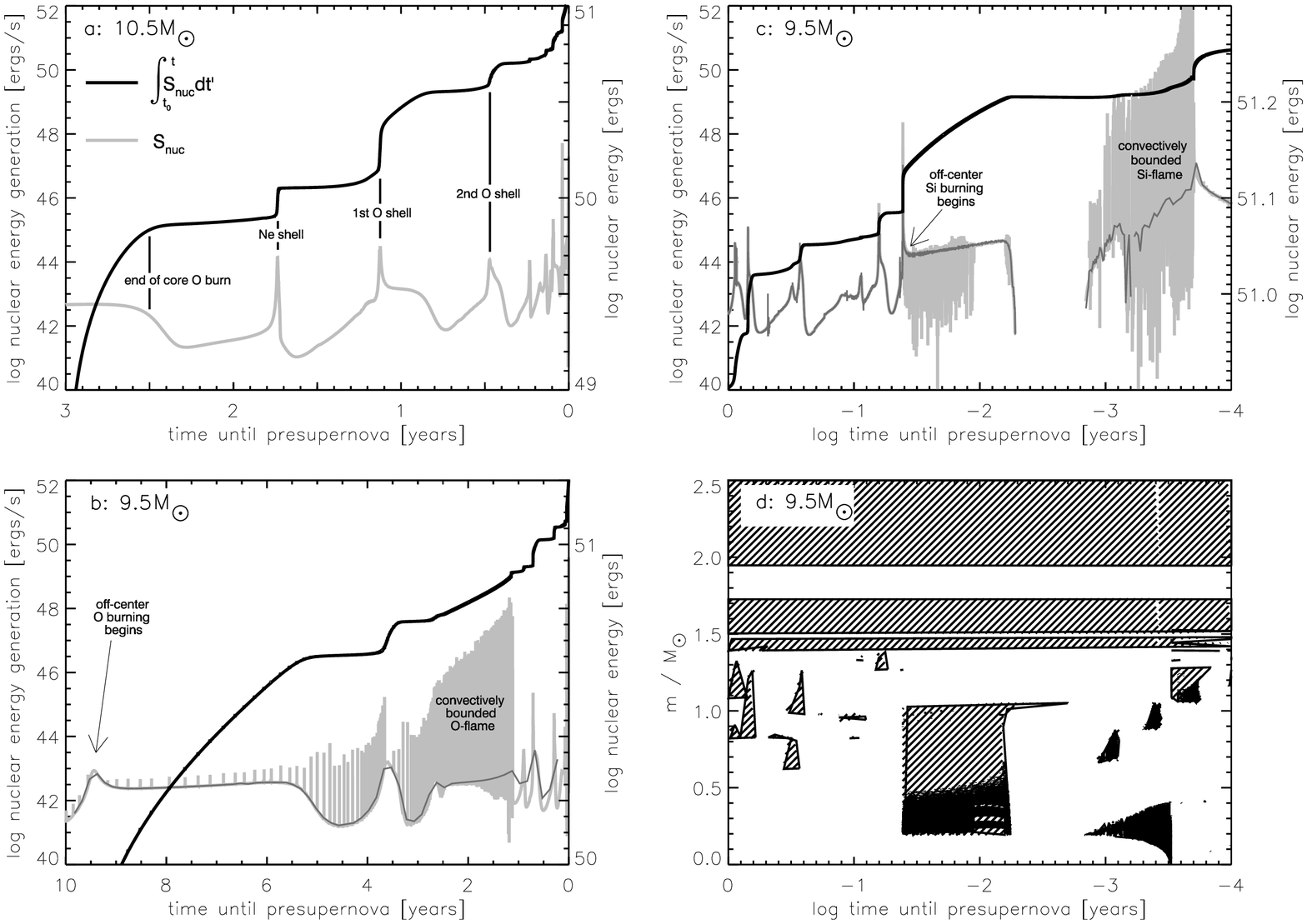}
\caption{Energy generation and convection in the 9.5 \Msun \ and 10.5
  \Msun \ models. a) Total nuclear energy generation for the 10.5
  \Msun \ model during the last 3 years of its life. The bottom line
  shows the nuclear energy generation rate and the top line, the
  cumulative integral of the energy generation rate as a function of
  time until the star's iron core collapses. The integral is
  initialized to zero at the beginning of the plot. Roughly
  10$^{51}$ erg is generated, mostly by off-center neon and oxygen
  shell burning. Central oxygen burning starts at 8.8 years before
  collapse (off-scale) and ends 2.3 y before collapse. b) A plot for
  the 9.5 \Msun \ model for the last 10 years of its life shows
  similar episodes of shell burning, but also an extended period
  during which a neon-oxygen burning CBF propagates to the center of
  the star (1 to 5 years before collapse). The many spikes are
  individual zones flashing (see text) and the solid line is the
  average. The density of flashing zones is greatest where the flame
  is progressing most rapidly A particularly strong oxygen shell
  burning episode happens 3.5 years before death, temporarily slowing
  the flame. c) The later evolution of the same 9.5 \Msun \ model on a
  logarithmic time scale shows further episodes of oxygen shell
  burning (1 to 10$^{-1.4}$ y) and pulsational silicon shell burning
  (10$^{-1.4}$ to 0.01 y before core collapse). Silicon burning
  ignites with a particularly violent flash ($\sim10^{48}$ erg
  s$^{-1}$) that almost becomes explosive. The solid line again
  indicates an average. The gap from log t = -2.3 to -2.8 is an
  interval of small negative energy generation. Starting about 13
  hours before death, a silicon burning CBF moves from 0.19 \Msun \ to
  the center, shortly after which the star's core collapses. d)
  Convective history during the same period shown in c).
  \lFig{flames}}
\end{figure*}

It is interesting to compare the properties of this oxygen burning
flame with those discussed in in \Sect{flame}. During the interval
from 3.0 to 1.1 years before collapse, the flame moves from 1300 km to
the stellar center while maintaining a flame temperature between 1.9
$\times 10^9$ K and $2.0 \times 10^9$ K. The average flame speed in
KEPLER was about 1 cm s$^{-1}$. This agrees reasonable well with the
the 0.5 - 2 cm s$^{-1}$ range given by \Eq{Timmes}.

After the oxygen CBF reaches the center of the 9.5 \Msun \ model,
there follows a period of about a year of residual oxygen shell
burning in the outer part of the star and Kelvin-Helmholtz contraction
in the inner part, during which the central density rises from $1.2
\times 10^8$ g cm$^{-3}$ to $8.6 \times 10^8$ g cm$^{-3}$.  Electron
capture in the oxygen convective shell and flame has already
substantially reduced $Y_e$ so that the core is more massive than the
cold Chandrasekhar limit\citep[][and \Fig{ye}]{Jon13}.  Eventually,
silicon ignites very degenerately, off-center at 0.186 \Msun. This
ignition is violent, but not quite hydrodynamic. The temperature at
the base of the silicon shell rises to $3.8 \times 10^9$ K at $2.9
\times 10^8$ g cm$^{-3}$ where $Y_e = 0.466$. The nuclear power
briefly reaches 10$^{48}$ erg s$^{-1}$ (log t = -1.4 in frame c of
\Fig{flames}), close to the value where convective transport would
break down (\Sect{convlim}).  This large power and rapid expansion of
the silicon core ($\sim 1$ km s$^{-1}$) launches a shock with speed
about 30 km s$^{-1}$ in the declining density gradient at the edge of
the core, but this shock attenuates as its momentum is shared with a
large amount of envelope mass and it never makes it to the stellar
surface before the core collapses.

Following the initial flash, silicon burns in a convective shell for
about two weeks (\Fig{flames}). For the first 8 days, the burning
occurs in a core that is still ``ringing'' from the strong
flash. During the last week the pulsations damp out though. Even
though there is no well-defined silicon burning CBF yet, the energy
generation during the pulsational phase varies wildly. Roughly $2
\times 10^{50}$ erg is released, and many strong sound waves are
launched into the envelope, but none are sufficiently violent
individually or collectively to eject the envelope.

After the silicon has burned to almost zero in the convective shell, a
silicon burning CBF begins its propagation to the center of the star.
This takes about 11 hours to move 600 km with an average flame speed
that, in KEPLER, was about 15 m s$^{-1}$. The higher speed here (as
compared with the oxygen-burning CBF) reflects the much higher
neutrino losses, 10$^{45}$ - 10$^{46}$ erg s$^{-1}$ that accompany the
greater burning temperature (\Fig{flames}), and to a lesser extent,
the smaller energy yield from burning silicon instead of oxygen. The
flame must advance faster to compensate (\Eq{flameeq};
\Sect{macroflame}). The speed is also not too far off from that
expected for a silicon CBF had one been included (\Sect{siflame}),
though the progression here is actually determined by the
Rayleigh-Taylor instability.

Once the silicon-burning CBF reaches the center, the inner 1.06 \Msun
\ is already composed of neutronized iron, but final collapse is
delayed until an additional 0.23 \Msun \ of silicon burns in a
shell. This takes 2.6 hours, and then the core collapses.

The 9.1, 9.2, 9.4, 9.6 and 9.7 \Msun \ models were qualitatively
similar to the 9.5 \Msun \ model, differing chiefly in the nature of
the silicon flash which occurred closer to the center of the star in
the heavier models. The 9.0 and 9.3 \Msun \ models, on the other hand,
were different and more similar to the silicon deflagrations to be
discussed in \Sect{siflash}. This because an additional oxygen burning
shell developed that allowed more cooling and greater degeneracy when
silicon ignited. The non-monotonic behavior of the late stages of
evolution due to the interplay of carbon and oxygen burning shells has
been discussed by \citet{Tug14}.

\subsection{Nuclear Power and Possible Mass Loss in the Last Decade}
\lSect{wind}

It has been suggested that the last years in a massive star's life
might be characterized a by very large mass loss rate driven by the
acoustic transport into the envelope of a portion of the vastly
super-Eddington powers developed in the star's convective shells
\cite[e.g.,][]{Qua12,Shi14}. The loss of even a fraction of a solar
mass during the last few years would substantially modify the spectrum
and light curve of the supernova that results when it dies.  Stars
around 10 \Msun \ are interesting in this regard because they spend a
longer time burning oxygen than their higher mass cousins. For
example, the 10.5 \Msun \ model begins oxygen burning 8.8 years before
iron core collapse. For a 25 \Msun \ star \citep{Woo07a}, the
corresponding time is 0.49 years.  The envelopes of these lower mass
stars are also very loosely bound, with net binding $\sim10^{47}$ erg,
and have steep density gradients around a degenerate core that could
serve to accelerate sound waves into shocks. Since the burning of a
solar mass of oxygen to iron releases about 10$^{51}$ erg, even
inefficient energy transport by sound waves could potentially have a
large effect.

\Fig{flames} shows the power developed by nuclear reactions in the
cores of 9.5 and 10.5 \Msun \ stars during their last 10 and 3 years
respectively. The 10.5 \Msun \ star is an example of a massive star
that ignites all burning stages in its center. It is extreme only in
being a light example of this class.  No CBFs form at any
point. The nuclear power after oxygen ignition stays at $\sim10^{43}$
erg s$^{-1}$ for a few months and then declines to around 1 to $2
\times 10^{42}$ erg s$^{-1}$ where it remains for 6 years until oxygen
is depleted in the stellar center. Over the next two years oxygen and
neon burn in shells, developing power that briefly climbs above
10$^{44}$ erg s$^{-1}$ when the shells ignite, and then, following a
brief phase of silicon burning, the iron core collapses to a neutron
star.

The 9.5 \Msun \ model, on the other hand, is characterized by CBFs
during both its oxygen and silicon burning phases. Like the 10.5 \Msun
\ model, it maintains a nuclear power over 10$^{42}$ erg s$^{-1}$ for
about a decade, but the mechanics of the burning is different. The
star ignites {\sl neon} burning 20 years before core collapse and, as
the neon depletes in the shell 10 years later, the burning transitions
into first a neon-burning, and then an oxygen-burning CBF. Since the
zones in the vicinity of the flame have nearly constant mass, the
spikes in energy generation from individual zone flashes indicate the
progress of the flame in \Fig{flames}.  Typical zoning in within 0.1
\Msun \ of the silicon flame was 10$^{30}$ gm (0.0005 \Msun) and the
full calculation of the 9.5 \Msun \ evolution required over 500,000
models. Much finer zoning would have been impractical and not added
greatly to our understanding. 3.5 years before core collapse, the CBF
pauses while oxygen burns in a growing convective shell. About
10$^{50}$ erg is released during this shell burning. Over the next 2.2
years, an oxygen-burning CBF moves to the center of the star
(\Sect{noexp}).

During its last year, the core experiences several episodes of oxygen
shell burning as the residual abundance of oxygen in the outer core
burns away. These are not CBFs, just regular shell burning, but they
help support the core while it cools by neutrino emission. Finally,
during the last two weeks, silicon ignites with a violent flash and
burns in a new CBF (silicon-burning). Two hours after that flame
reaches the center, the iron core collapses. Large powers (over
10$^{46}$ erg s$^{-1}$) are developed during and after the
silicon-burning CBF propagation, but this energy is deposited so late
that, baring a hydrodynamical event, it probably has little effect on
presupernova mass loss.
 
Without multi-dimensional simulation the real physical nature of these
oxygen and silicon burning CBFs is difficult to know. The sputtering
of individial zones is artificial, but still the matter in the flame
is degenerate. Localized runaways may result in the more effective
production of acoustic energy than ordinary convection.

\subsection{The Limits of Convection}
\lSect{convlim}

In the $9.5\,\Msun$, $9.6\,\Msun$, and $9.7\,\Msun$ models, the
maximum luminosity in the silicon convective shell during silicon
ignition briefly reached $1.2$, $0.8$, and
$0.7\times10^{48}\,$erg$\,$s$^{-1}$, respectively, and the expansion
speed exceeded several km$\,$s$^{-1}$.  The silicon burning luminosity
peaks as the burning density at the shell declined below about
$2\times10^8\,$g$\,$cm$^{-3}$ when the burning temperature was near
$3.8 \times 10^9\,$K. As will now be shown, luminosities this large are
close to the limit of what can be carried by convection at these radii
and densities.

The maximum power that convection can carry is approximately
\begin{equation}
L_{\rm max} \ \approx \ 4 \pi r^2 \rho v_{\rm conv} f C_\mathrm{P} T,
\end{equation}
where $r$ is the radius of the shell, about $500\,$km; $\rho$ is the
density when the luminosity is maximal, about
$2\times10^8\,$g$\,$cm$^{-1}$; $v_\mathrm{conv}$ is the convection
speed which must be substantially subsonic, i.e., of order
$1000\,$km$\,$s$^{-1}$; $C_\mathrm{P} T$ is the heat content, close to
10$^{17}\,$erg$\,$g$^{-1}$; and $f \ll 1$ reflects the fact that the
luminosity can only remove a small fraction of the heat content of a
zone in a convective crossing time without shutting off the
convection.  For $f \sim 0.1$, $L_{\rm max} \sim
10^{49}\,$erg$\,$s$^{-1}$.  This limit is consistent with what was
observed in KEPLER. Models that developed greater luminosities were
unable to transport the power by ordinary convection and a localized
runaway developed.

\section{EXPLOSIVE SILICON BURNING - $9.8\,\Msun$ - $10.3\,\Msun$}
\lSect{siflash}

From $9.8\,\Msun$ through $10.3\,\Msun$ and also for $9.0\,\Msun$ and
$9.3\,\Msun$ (8 models), oxygen burning, ignites off center, and then
burns to the center of the star in a CBF. Subsequently, a series of
oxygen shell burning episodes culminate in the production of a roughly
Chandrasekhar-mass of degenerate silicon.  Substantial neutronization
has already occurred by this point resulting in a decrease in the
Chandrasekhar mass before silicon ignites (\Fig{ye}). Further neutrino
cooling after oxygen depletion thus lead to increased degeneracy and a
potentially explosive configuration.

Silicon thus ignites in these stars as a powerful deflagration
generating a shock wave that propagates into the hydrogen envelope.
While the calculations unambiguously show the occurence of some sort
of dynamic event, accurate results are difficult to obtain because,
once the runaway becomes localized to an off-center point, the further
evolution is inherently three dimensional.

\subsection{Evolution at $10\, \Msun$}
\lSect{10sun}

For illustration, consider the $10.0\,\Msun$ model.  This model begins
its runaway in a fashion similar to the $9.5\,\Msun$ (top frame of
\Fig{comp9.5}), but nuclear energy generation rapidly grows to
super-critical levels (\Sect{convlim}).  Igniting at $0.0203\,\Msun$
(instead of $0.201\,\Msun$ for the $9.5\,\Msun$ model) and a density
$4.97 \times 10^8\,$g$\,$cm$^{-3}$, the temperature quickly rises from
$3.23 \times 10^9\,$K at ignition (when convection first begins),
reaching $4.2 \times 10^9\,$K just as the maximum luminosity in the
shell exceeds $10^{49}\,$erg$\,$s$^{-1}$.  This temperature, $T_9
\approx 4.2$, plays a similar role to $T_9 = 0.8$ in the carbon
runaway in SN Ia \citep{Woo04b}.  Silicon flashes that do
not reach this threshold temperature do not become explosive and cause
hydrodynamical mass ejection (\Sect{noexp}).  Once the temperature
exceeds this value, however, convection freezes out and a single zone
runs away in isolation, reaching a maximum of $6.4\times 10^9\,$K only
1.7 ms later.  During this time, the silicon and other intermediate
mass elements in that zone burn to nuclear statistical equilibrium.

At this point, a localized flame has been born.  What happens next is
uncertain.  A density inversion might develop leading to
Rayleigh-Taylor instability, as in a SN Ia, but this was not
immediately apparent in the KEPLER model after the runaway of a single
zone. If convection was turned off when the luminosity reached
$10^{49}\,$erg$\,$s$^{-1}$, the burning still propagated for brief
period as a detonation wave through about $0.01\,\Msun$ \ (6
zones). This happened because the phase velocity for the burning
implied by the adiabatic temperature gradient at such high
luminosities was supersonic.  Had convection been turned off at a
lower luminosity this supersonic burning probably would not have
occurred. The fact that a detonation induced by a shallow temperature
gradient failed to survive is also suggestive that detonation may be
difficult to achieve in a silicon-rich composition.

Once the detonation died, expansion of these burned layers \emph{did}
result in a density inversion of about $15\,\%$ - comparable to the
decrement seen in carbon-deflagration models for SN Ia.  So the likely
outcome here is a silicon deflagration - but in how extensive a
region?  How much mass would burn?  Based upon the well-studied case
of off-center ignition in SN Ia, the runaway might initially ignite at
a single point \citep{Non12}, not in a symmetric spherical shell.  If
only a single point ignites and gives rise to a single plume of
burning, the total mass consumed in the deflagration may be small
\citep{Mal14}.  Until the core expands significantly, degenerate
burning would continue however, possibly igniting in more than one
location.  Without detailed 3D studies \citep[for an analogue
  see][]{Zin13}, it is difficult to say just how much silicon burns.
If a subsequent transition to another detonation occurs later, then an
appreciable fraction of the entire core could burn, but this may be
even more difficult to achieve for silicon than for carbon and oxygen.

Given the smaller energy yield from silicon burning (as opposed to
carbon burning) to iron, the silicon core will not be fully disrupted,
even if a significant fraction burns.  The binding energy of the star
at this point is $4.8 \times 10^{50}\,$erg.  Silicon burning to iron
releases only $4.9 \times 10^{50}$ erg for each solar mass of silicon
burned.  The core is relativistically degenerate, however, with a
structural adiabatic index close to $4/3$, so even a little burning
causes significant expansion. If that expansion is sufficiently rapid,
the shock wave it creates can eject the loosely bound envelope.
Typically the net binding energy of most of the hydrogen envelope is
near $10^{47}\,$erg (\Tab{endstate}).  For the $9.8\,\Msun$ model at
onset of the silicon flash, the total star mass is $9.48\,\Msun$ and
the binding energy outside of the inner $2.0\,\Msun$ is $1.9 \times
10^{48}\,$erg.  Outside of the inner $2.3\,\Msun$, it is $1.9 \times
10^{47}\,$erg. Burning $0.01\,\Msun$ to iron yields $4.8 \times
10^{48}\,$erg, so there is no dearth of energy to cause an explosion
of some sort.  Transporting this energy out of the core by a shock
wave though, and giving the envelope enough \emph{momentum} to exceed
its escape speed is much more difficult.

Other models in the $9.8\,\Msun$ though $10.3\,\Msun$ range
experienced a similar evolution to $10.0\,\Msun$, though the heavier
stars ignited silicon closer to or at the center (\Tab{siflash}).  The
composition of the $10\,\Msun$ model at the time silicon ignites is
given in \Fig{comp10}.  Above $10.3\,\Msun$, silicon ignited gently and
the stars evolved to iron core collapse with their envelopes still
intact.  The $10.4\,\Msun$, $10.5\,\Msun$ and $11.0\,\Msun$ models
ignited all advanced stages, carbon, oxygen, and silicon burning at
center of star and can thus be considered ``normal'' core-collapse
supernovae.

% fig 9 - composition, T and density for 10 Msun at si ignition
\begin{figure}
\begin{center}
\includegraphics[width=0.475\textwidth]{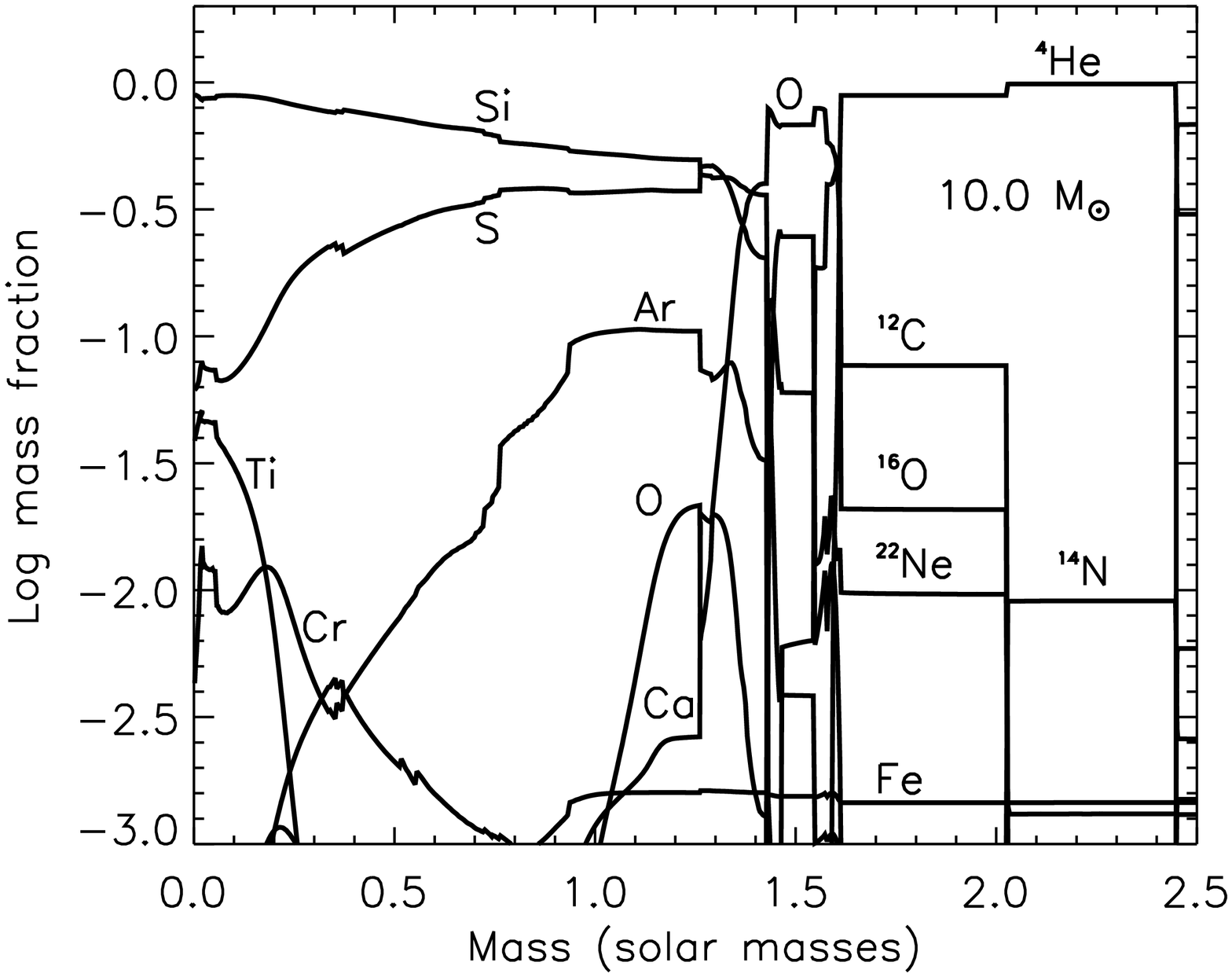}
\hfill
\includegraphics[width=0.475\textwidth]{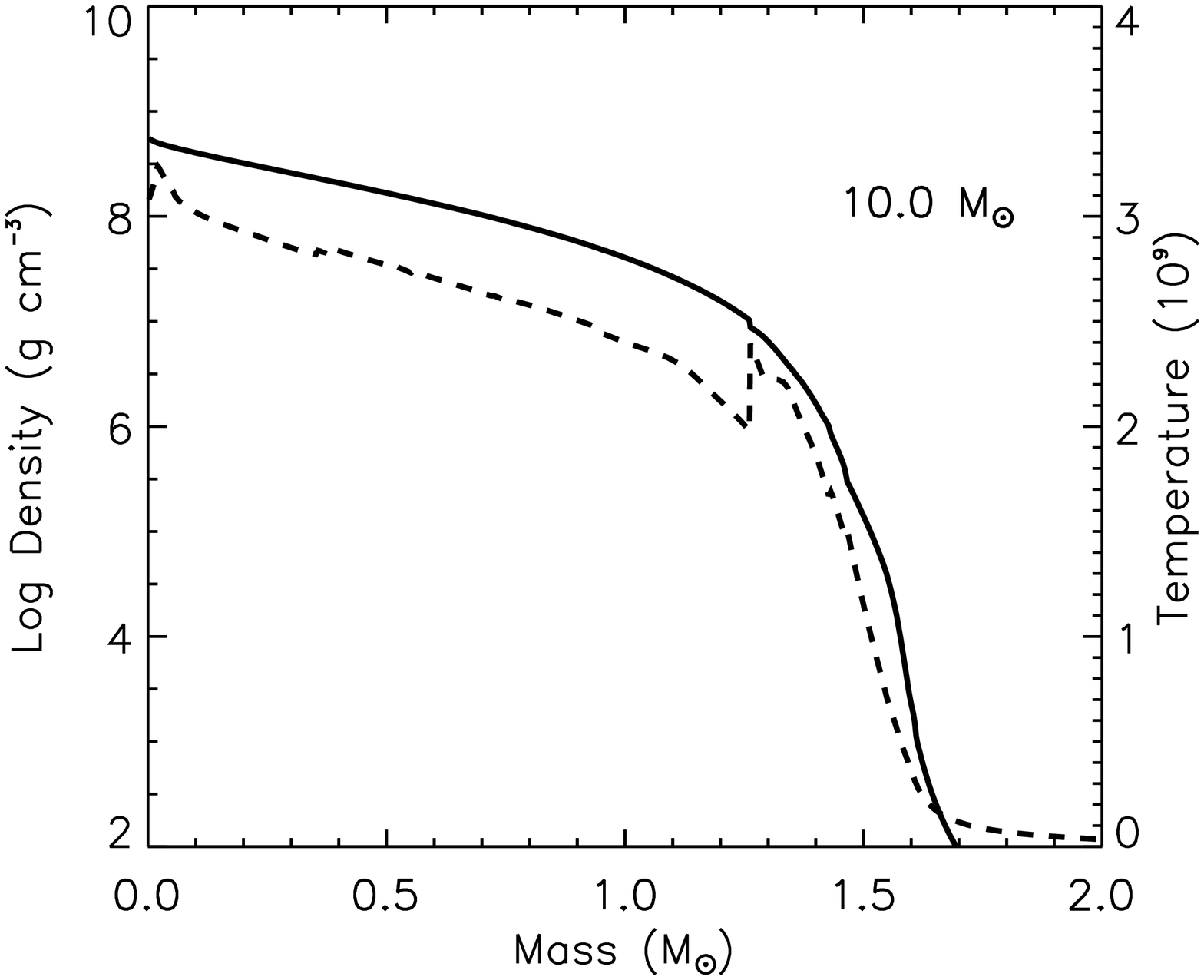}
\caption{{\sl Top:} Composition of the $10\,\Msun$ model at the time
  silicon ignites with a violent flash at $0.02\,\Msun$.  Only the
  inner $2.5\,\Msun$ is shown and the helium core mass is
  $2.45\,\Msun$. {\sl Bottom:} Temperature (dashed) and density (solid
  line) structure at the same time. The central density and
  temperature are $5.51 \times 10^8\,$g$\,$cm$^{-3}$ and
  $3.08\times10^9\,$K, respectively.  In the location of silicon
  ignition the density and temperature are
  $4.97\times10^8\,$g$\,$cm$^{-3}$ and $3.23\times 10^9\,$K
  respectively (\Tab{siflash}).
\lFig{comp10}}
\end{center}
\end{figure}

\subsection{Mass Ejection}
\lSect{meject}

The observable outcome of these models depends upon the amount of
silicon that burns hydrodynamically.  A small amount of burning does
not eject the envelope, or ejects it so slowly and so close to the
time of iron core collapse that there is little observable distinction
compared to a star in which no silicon flash occurred.  Because of the
uncertainties inherent in a 1D simulation a variety of outcomes was
explored and parametrized by how much silicon burns explosively.
\Tab{outcome} gives, for each case where explosive silicon burning
occurred, the amount of silicon burned in the initial flash
(``Fe-Mass''), the delay time between the silicon flash and iron-core
collapse, the photospheric radius of the presupernova star (which may
be small if the ejected envelope has already recombined), the velocity
of the ejecta from the silicon flash (if any) at the time of iron core
collapse, and the luminosity of the star or remaining core when the
iron core collapses.  Cases of high velocity and large radius
($\sim10^{15}\,$cm) or high luminosity indicate a prior supernova in
progress when the core collapses.

In all cases, the burning was allowed to propagate artificially, for a
time, at nearly sonic speed by leaving convection on well past the
stage where the luminosity exceeded $10^{49}\,$erg$\,$s$^{-1}$.  Some
variation in outcome was achieved by changing the efficiency of the
convection using a multiplier on the calculated convective velocity
(``Conv. Param.'' in \Tab{siflash}).  No physical significance is
attributed to this operation.  It is just a way of producing variable
amounts of burning.  The upper bound to the mass that burns, around
$0.75\,\Msun$, occurs when the density becomes sufficiently low that
the complete burning of silicon to iron produces too little energy to
raise the fuel to a temperature where it world burn in a Courant time
scale (sound crossing time for a zone). Thus a detonation wave, if
there were one, would likely stall after burning this mass.

Having explosively produced the given ``Fe Mass'', the core expands
sending out a strong pulse that steepens into a shock wave in the
steep density gradient at its edge. Afterwards, the core pulses a few
more times and experiences a Kelvin-Helmholtz phase of variable
duration before finally settling down to ignite silicon shell burning
stably.  Continued evolution in all cases gave an iron core that
collapsed to a neutron star, presumably launching some sort of
additional explosion (\Sect{rotate}).  The delay time between the
onset of the first explosive silicon flash and the final iron core
collapse is given as ``Delay Time'' in \Tab{siflash} and varied from a
couple of weeks to a couple of years.  The final velocity structure
for several $10\,\Msun$ models parametrized by the amount of silicon
that burned is given in \Fig{un10}.

% fig 10 - velocities 10.0
\begin{figure}
\includegraphics[width=\columnwidth]{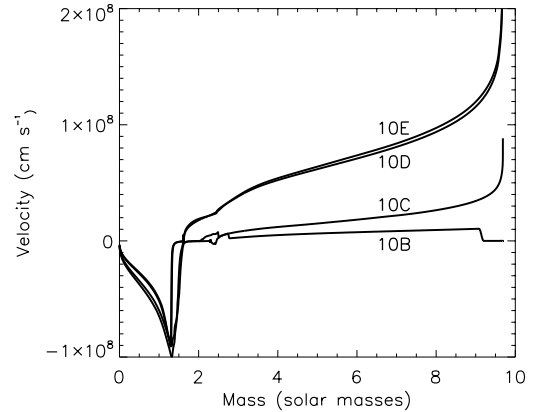}
\caption{Velocities in the $10.0\,\Msun$ model at the time of iron
  core collapse.  The models are defined in \Tab{outcome} which also
  gives the delay time for each model between the silicon flash and
  iron core collapse (``Delay Time''). Burning more silicon in the
  flash gives more kinetic energy to the envelope and increases the
  delay time.  Models~10C, 10D, and 10E have envelopes that have
  already expanded an appreciable distance, whereas in Model~10B, the
  shock is still in the envelope when the iron core
  collapses. \lFig{un10}}
\end{figure}

In some cases, when only a little silicon burns, like in Models 10A
and 10B, the shock wave either never makes it, or barely makes it to
the surface before the iron core finishes evolving and collapses to a
neutron star. Thus the ``presupernova'' radius and main light curve
are not appreciably altered. There might be appreciable changes,
however, in the supernova luminosity at break out and in the
presupernova density structure near the iron core.

In other cases where the amount of silicon burned was greater,
however, envelope ejection occurs well before core collapse (Models
10C, 10D, 10E) and the observable supernova was very different.  Essentially
two supernovae occurred in rapid succession, and the second one could be
very bright.  In these cases the velocity of the outer layers was large
and, if enough time elapsed for significant recombination, the
photospheric radius and radius of the outermost eject may differ -
hence the two entries for the photospheric radius and the ``edge'' in
\Tab{outcome}.

\section{SUPERNOVAE, BRIGHT AND FAINT, SINGLE AND DOUBLE}
\lSect{ultra}

\subsection{Rotation, Energetics and Light Curves}
\lSect{rotate}

Numerous studies \citep[e.g.,][]{Woo92,Fry99,Des06,Des07} have shown
that (at least) a low energy explosion, $\sim10^{50}\,$erg with
$\sim0.01\,\Msun$ of $^{56}$Ni ejected, is an inevitable consequence
of the accretion-induced collapse of a white dwarf.  If nothing else,
the neutrino-powered wind that accompanies neutron star formation will
release that much energy \citep{Qia96,Kit06}.  The cores in
\Fig{presndn} are so hydrodynamically detached from their low density
envelopes, and their mantles of helium and carbon so small, that their
collapse will closely resemble bare white dwarfs.  Calculations of
accretion-induced collapse are thus directly applicable to the study
of supernovae in this mass range and similar results are expected and
have been obtained for non-rotating stars by \citet{Kit06} and
\citet{Bur07}.

With rapid rotation and large field strengths, both the energy and
mass ejected could be considerably greater \citep{Des07}, but the
rotation rates and field strengths required may be unrealistically
large.  Given the long period spent by single stars as a red
supergiant, magnetic torques will likely lead to the core rotating
slowly at the time of its collapse.  A single test case, a
$10\,\Msun$, solar metallicity model evolved here to core collapse
including rotation and magnetic torques as in \citet{Heg05}, had an
angular momentum interior to the base of the oxygen shell at $1.46\,
\Msun$ of $5.7 \times 10^{47}$ erg s. The angular momentum in the iron
core alone, $1.33\, \Msun$, was $5.0 \times 10^{47}$ erg
s$^{-1}$. Assuming a pulsar moment of inertia, $I \approx 0.36 MR^2$
\citep{Lat89,Lat01} and radius 12 km, this implies a pulsar period
near 17 ms and a gravitational mass of 1.2 to $1.3\, \Msun$. This
estimated period compares quite favorably with observationally
inferred estimates for the Crab pulsar rotational period at birth of
$\sim21\,$ms \citep[e.g.,][]{Mus96,Heg05}.  Such a slow period for a
neutron star implies a rotational kinetic energy of less than
$10^{50}\,$erg, so even the inclusion of rotation is unlikely to
produce an explosion of $\sim10^{51}\,$erg.

We note in passing that the low explosion energy, essentially from a
spherically symmetric neutrino-powered wind, and the lack of
appreciable accretion during the explosion, might imply a low
``kick-velocity'' for the Crab pulsar \citep{Won13}.  Measurements
\citep{Kap08} suggest that the Crab pulsar moves slower than most
others, and even that speed could be a consequence of a progenitor
that had run away from a binary and a prior explosion.

For now, to illustrate the qualitative features of the resultant
supernovae, we adopt here an explosion energy at the time of iron core
collapse of $2 \times 10^{50}\,$erg, and explore the consequences.
Since the interaction with the ejected envelope considerably slows the
ejecta from the core collapse and can even radiate a large part of the
energy as light, this energy must be evaluated somewhat differently
than the customary ``kinetic energy at infinity'' criterion.  Here we
use the explosion energy to be defined as the net energy on the grid
(gravitational binding energy plus internal energy) when all nuclear
reactions have stopped and the explosion is well underway.  This is,
however, well before any shock has erupted from the surface of the
star or any bright display commenced.

Using a piston to impart a net energy of $2.2 \times 10^{50}\,$erg to
the matter outside the collapsed core in Model~10B, the resulting
shock from core collapse overtook the shock from the earlier silicon
flash just as the latter was breaking out (\Fig{10blite}).  The
resulting light curve is thus very similar to what would be obtained
had there been no silicon flash - a relatively faint, long lasting
Type IIp supernova.  This solution should be representative for all
the models in \Tab{outcome} where the velocity at the ``edge'' of the
star is zero and where the photospheric radius approximately equals
the edge radius, i.e., the star has not significantly expanded before
the iron core collapsed.

% fig 11 - velocities and light curve 10.0b
\begin{figure}
\begin{center}
\includegraphics[width=0.475\textwidth]{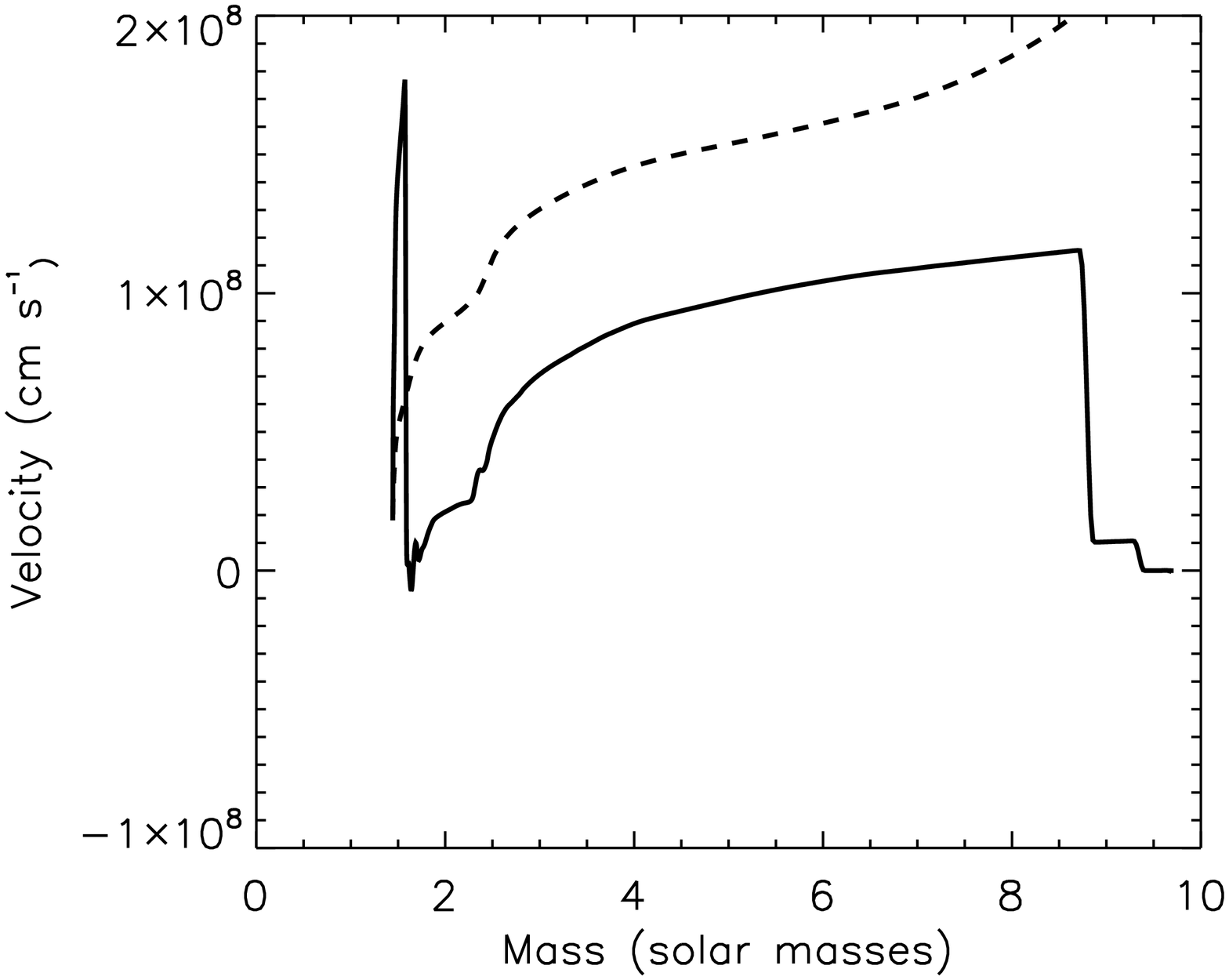}
\hfill
\includegraphics[width=0.475\textwidth]{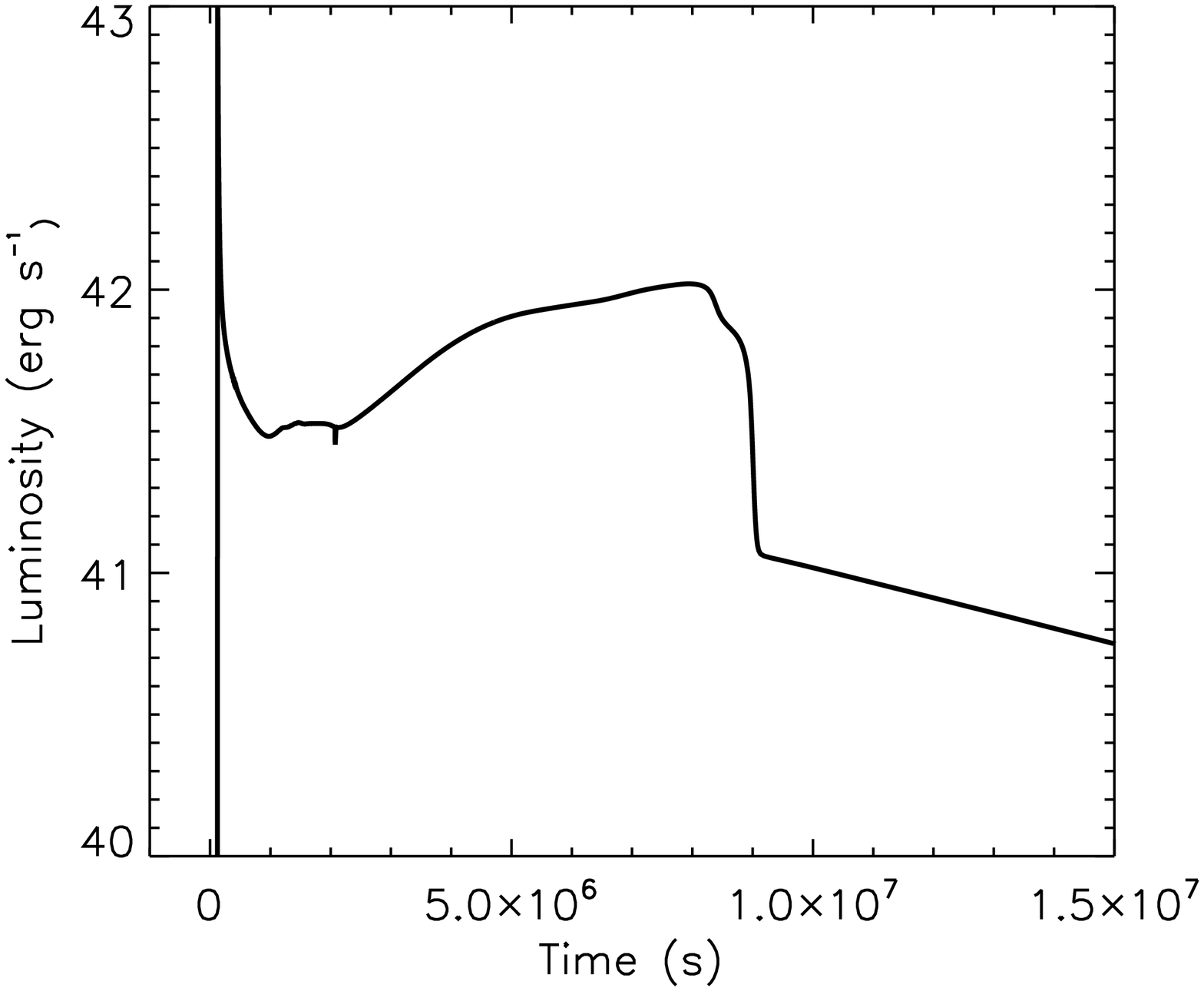}
\caption{{\sl Top:} Velocity of Model~10B shortly before the break
  out of the main shock (solid line; $9.5 \times 10^4\,$s after iron
  core collapse) and after the ejecta have achieved their terminal
  speed (dashed line; $2.0 \times 10^7\,$s after core collapse).  Note
  the ``bump'' at $9.31\,\Msun$ with a speed of $100\,$km$\,$s$^{-1}$.
  This is the leading edge of the shock previously launched by the
  silicon flash in this model (\Fig{un10}).  Also visible in the early
  velocity curve (solid line) is a spike in velocity near the origin.
  This is the location of the reverse shock from the primary
  explosion.  At late times the velocity structure is not affected by
  the small pulse that proceeded it. {\sl Bottom:} Bolometric light
  curve of Model~10B.  The low kinetic energy of the explosion, $2.2
  \times 10^{50}\,$erg, gives rise to a single faint Type IIp
  supernova.  Peak luminosity at shock break out is off scale, but
  equals $1.2 \times 10^{44}\,$erg$\,$s$^{-1}$.  At late times the
  light curve is powered by the decay of $0.022\,\Msun$ of
  $^{56}$Co.\lFig{10blite}}
\end{center}
\end{figure}

Very different results are found for Model~10D and similar models
where a large fraction of the silicon core burns.  In these cases the
silicon flash promptly ejects the entire hydrogen envelope with speeds
$\sim1000\,$km$\,$s$^{-1}$.  Equally important, the strong flash
causes the core to expand to such low density that the
Kelvin-Helmholtz time scale for recontracting and reigniting silicon
burning becomes very long - months to years rather than days.  As a
result, significant mass moves to 10$^{14}\,$cm - 10$^{15}\,$cm before
the iron core collapses.  The collision of shells at such large radii,
where the matter is almost optically thin, converts streaming kinetic
energy into light with unusually high efficiency.  A substantial
fraction of the entire explosion energy, $\sim10^{50}\,$erg, comes out
as light.  Also because of the long delay, there is time for two
supernova-like displays from a single star's death.

The first is a relatively faint explosion,
$\sim10^{41}\,$erg$\,$s$^{-1}$ (\Fig{10dlite}) as the silicon flash
expels the red giant envelope.  The total kinetic energy associated
with event 1 in the figure is only $4.7 \times 10^{49}\,$erg (see also
\Fig{un10}). This event has faded away about $9$ months ($2.4 \times
10^7\,$s) before the onset of the second much brighter supernova.  Had
the silicon flash energy been less, a single supernova with unusual
time structure - like the superposition of the two figures - would
have resulted.  Similar low energy Type II supernovae have been
considered by \citet{Lov13}, albeit with a different central energy
source.

% fig 12 - light curve 10.0d
\begin{figure}
\begin{center}
\includegraphics[width=\columnwidth]{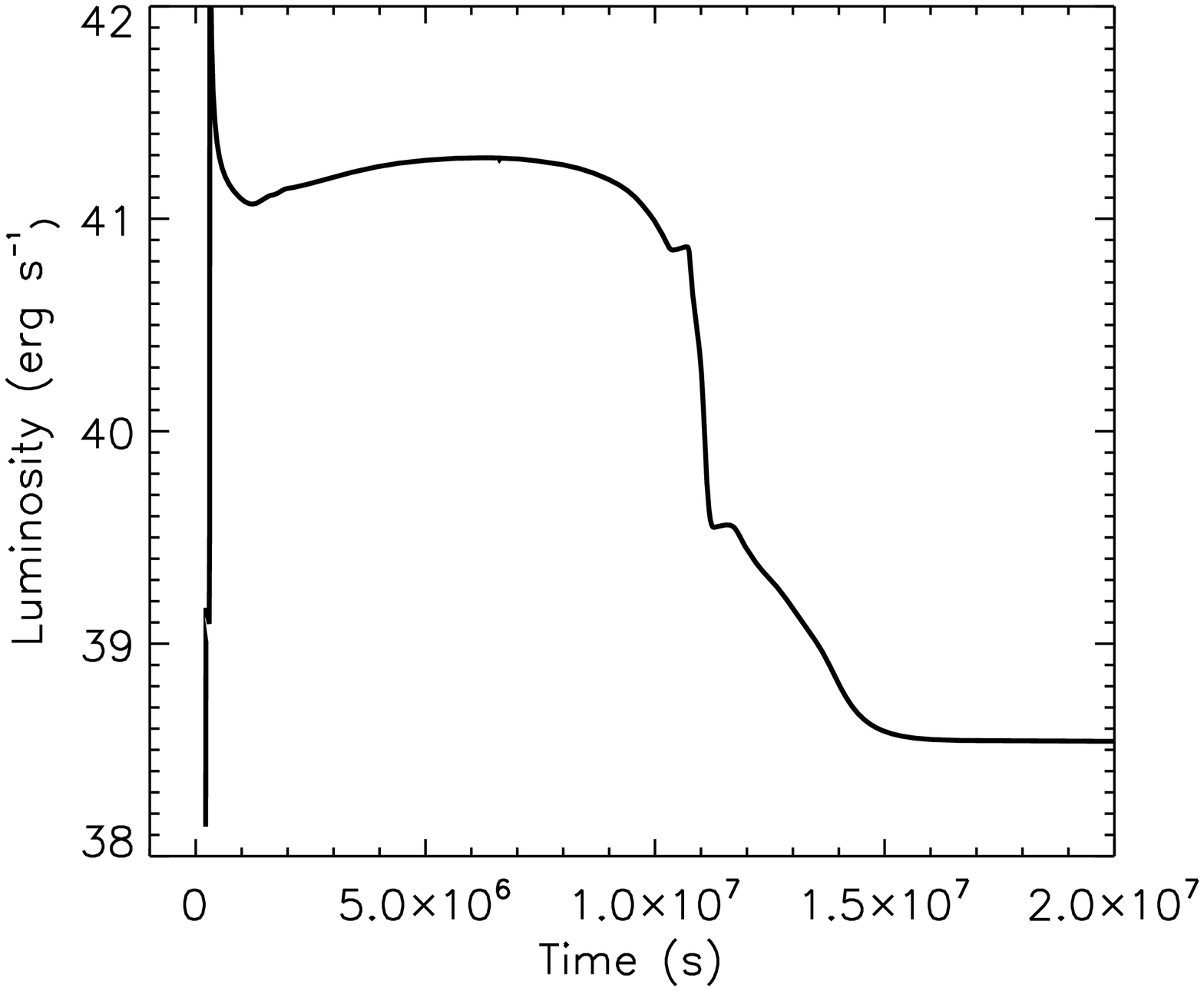}
\hfill
\includegraphics[width=\columnwidth]{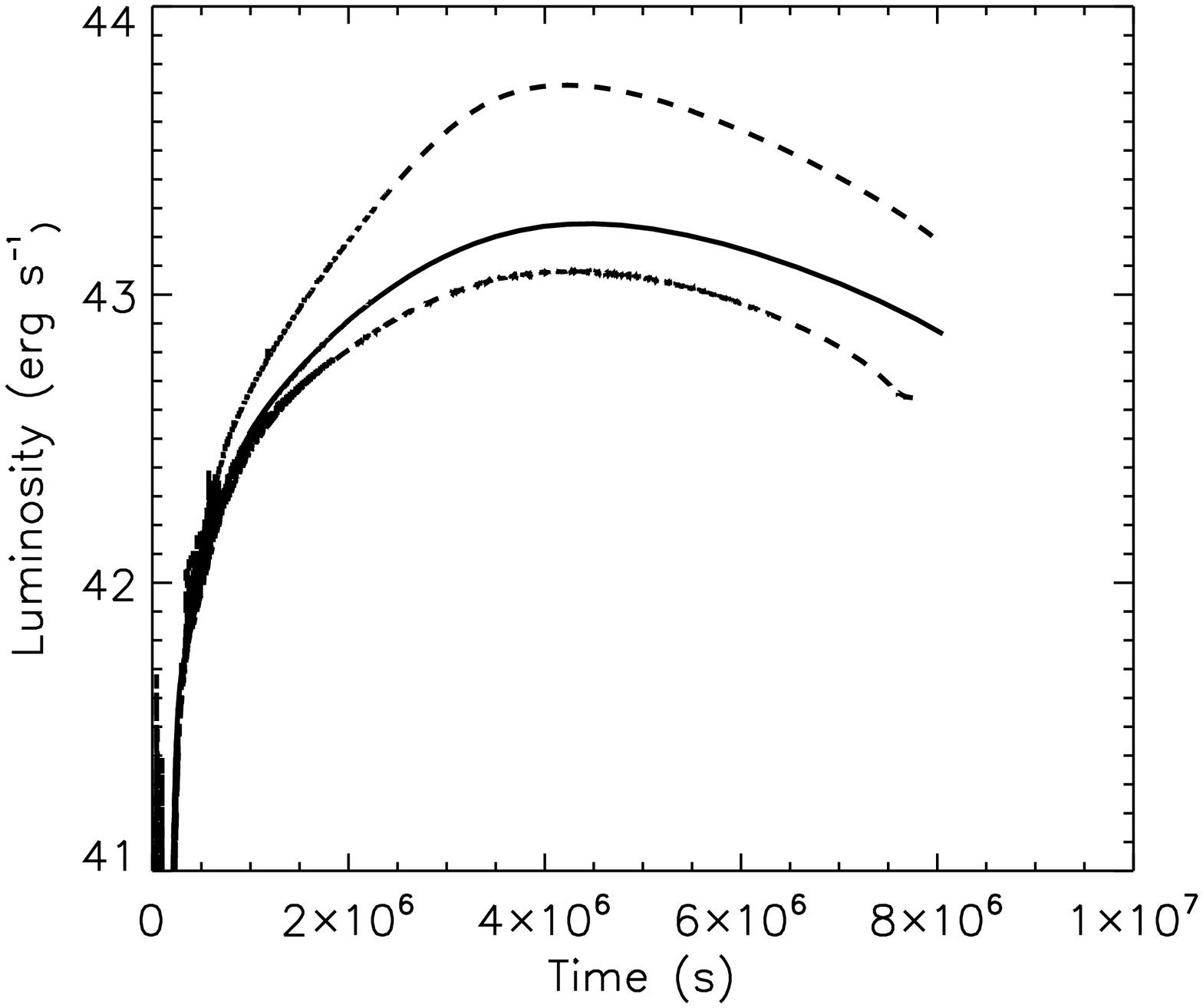}
\caption{Light curve of Model~10D. {\sl Top:} The silicon shell flash
  produces a faint Type IIp supernova which begins $400$ days before
  iron-core collapse produces a second supernova.  Peak luminosity at
  break out for this first event is $\sim3\times10^{43}\,$erg$\,$s.  A
  typical temperature on the plateau is $6200\,$K. {\sl Bottom:} A
  second, much brighter supernova is produced when the more energetic
  explosion produced by core collapse runs into the envelope at
  $\sim10^{15}\,$cm.  The solid line is for an explosion energy of
  $2.0\times10^{50}\,$erg.  Dashed lines are for $1.4$ and
  $4.5\times10^{50}\,$erg.  Zero time in the bottom frame corresponds
  to $3.47\times10^7\,$s in the top frame.  Note the two order of
  magnitude increase in the luminosity scale.  As the fast moving
  ejecta snowplows into the envelope, a large spike in density is
  created at the interface.  The geometrically thin nature of this
  spike, which is unphysical when considered in 2D or 3D, causes
  numerical difficulties that precluded running the light curve beyond
  $8\times10^6\,$s.  \lFig{10dlite}}
\end{center}
\end{figure}

The second supernova is very bright.  If the envelope has moved an
optimal distance, $\sim10^{15}\,$cm, before being overtaken by the
shock generated by core collapse, kinetic energy can be converted into
light with high efficiency. Interestingly, the energy from the silicon
flash in many models in \Tab{outcome} gives the right value to make
bright displays from the low energy explosion of stars of only
moderate mass.  Once the shells have been launched and collide, the
outcome resembles what has been proposed for pulsational-pair
instability supernovae at much higher mass \citep{Woo07a} and poses
some of the same computational challenges.  In particular, when
calculated in 1D, the colliding shells produce a very dense,
geometrically thin shell in which most of the mass resides
(\Fig{10dspike}).  This shell poses computational difficulty because
many Lagrangian shells have essentially the same radius and is also
unphysical.  In reality, the collision will produce mixing that must
be studied in at least two dimensions \citep{Che14}.  Qualitatively,
the bolometric light curves shown in \Fig{10dlite} will probably not
change appreciably (up to the point where they could be calculated),
but the composition will be mixed and the spectrum altered.

% fig 13 - density spike 10.0d
\begin{figure}
\includegraphics[width=\columnwidth]{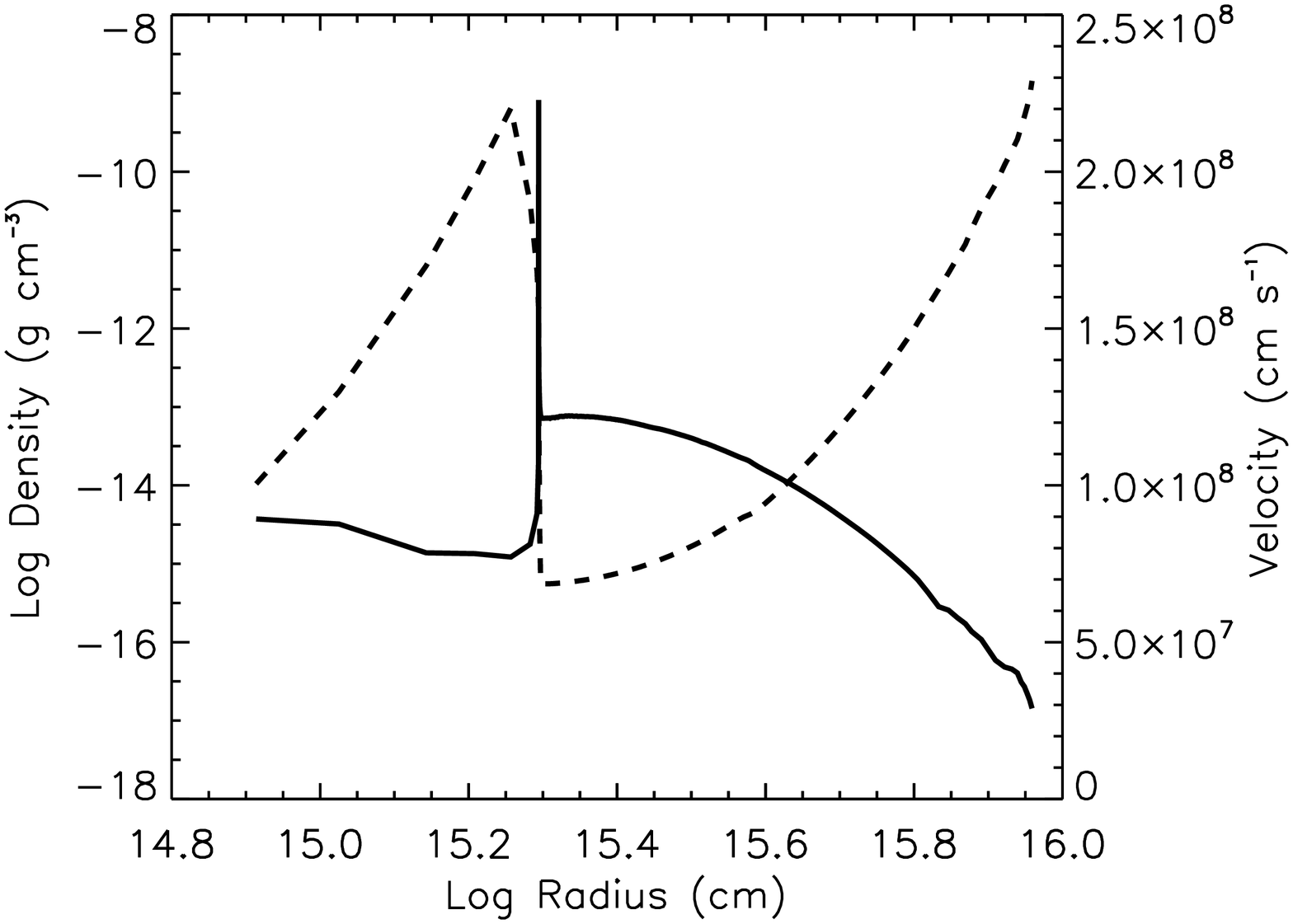}
\caption{Velocity (dashed line) and density (solid line) in Model~10D
  at $8\times10^6\,$s for the $2.0 \times 10^{50}\,$erg
  explosion. \lFig{10dspike}}
\end{figure}

\subsection{Nucleosynthesis}

Stars in this mass range are not prolific sources of common elements
\citep{Hil84,May88}.  The steep density gradient around the collapsing
core implies a low shock temperature in all but the matter very
nearby, so explosive nucleosynthesis is minimal.  The compact
structure of the presupernova core also implies thin carbon and oxygen
shells, so pre-explosive nucleosynthesis is small as well.  Indeed,
the chief nucleosynthetic contribution of stars in this mass range may
be part of the \textsl{s}-process, for those stars that develop a thin
helium shell and pass through an asymptotic giant branch phase ($M <
9.00\,\Msun$ here), or the \textsl{r}-process from the neutrino
powered wind when the neutron star is born.

The synthesis of intermediate mass elements and iron is nevertheless
interesting for predicting the sorts of abundances that might be found
in the remnants of supernovae in this mass range, and for the effect
of $^{56}$Ni on the light curve. The nucleosynthesis for the two
representative cases in which explosions were simulated, Models 10B
and 10D, is summarized in \Tab{nucleo}.  The synthesis of $^{56}$Ni is
quite sensitive to the location of the piston used to drive the
explosion and will remain uncertain until a real (i.e.,
multi-dimensional) explosion model can be computed.  For Model~10B,
two locations for the piston were explored.  One, at $1.301\,\Msun$,
is the traditional location used in many other studies, the base of
the oxygen shell where the entropy experiences an abrupt rise above
$S/N_\mathrm{A}k_\mathrm{B}=4.0$.  Given the sharp density decline
there, very little $^{56}$Ni is made.  The other location,
$1.245\,\Msun$, at the edge of the neutronized iron, was the deepest
location likely to be ejected in any model. This is in part because
calculations of neutrino-transport models show that the iron core is
generally a lower bound to the mass incorporated into the bound
baryonic remnant, but also because ejection of even a few hundredths
of a solar mass of such neutron-rich matter would greatly overproduce
rare nuclei in the iron group.
  
In any case, the ejection of more iron-group matter would not have
increased the $^{56}$Ni yield.  Model~10D had a piston situated at the
$S/N_\mathrm{A} k_\mathrm{B} = 4.0$ point, but the density decline
there was not so steep since the weaker silicon flash had not ejected
so much matter and there had been less time to cool and contract.  In
summary, it seems that $^{56}$Ni production will likely be in the
range $0.01\,\Msun$ to $0.04\,\Msun$.  The production of $\sim0.01
\Msun$ of $^{56}$Ni in $10\,\Msun$ explosions is consistent with
previous studies \citep{Kit06,Wan09}, and pending further
multi-dimensional modeling \citep[e.g.][]{Mel15} and studies of the
nucleosynthesis in the neutrino-powered wind, we believe it to be a
good factor of two estimate.  Lighter elements are less sensitive to
the simulation of the explosion and therefore are more accurately
determined.

\section{CONCLUSIONS}
\lSect{conclude}

The presupernova evolution of stars in the $6.5\,\Msun$ to
$13.5\,\Msun$ range has been explored with emphasis upon stars from
9.0 to $10.3\, \Msun$.  These are stars that ignite oxygen off center.
An important component of these studies is the use of a large network,
including the necessary weak interactions for altering the electron
mole number, $Y_\mathrm{e}$, during all stages of the post-helium
burning evolution \citep{Jon13}.  Particularly important is the
neutronization that goes on during oxygen shell burning and decreases
the effective Chandrasekhar mass of the core, making it prone to
collapse in the absence of strong burning shells. Also important and
novel is our treatment of the propagation of the oxygen and silicon
burning CBFs, especially using a subgrid model to describe the oxygen
CBF propagation.

We find that evolution in this mass range can be categorized by five
possible outcomes (\Fig{fates}). In order of increasing mass these
are: \textit{1)} carbon-oxygen white dwarfs (\Sect{CO}; below 7.0
$\Msun$); \textit{2)} neon-oxygen white dwarfs or electron-capture
supernovae (depending upon uncertain mass loss rates; \Sect{NeO}; 7.0
- 9.0 $\Msun$); \textit{3)} stars that ignite degenerate silicon
burning off-center in a strong flash, but which remain
hydrodynamically stable until iron-core collapse (\Sect{noexp}; 9.1,
9.2, and 9.4 - 9.7 $\Msun$); \textit{4)} stars for which the silicon
flash is so violent as to lead to a localized deflagration and
possible envelope ejection (\Sect{siflash}; 9.0, 9.3, and 9.8 - 10.3
$\Msun$); and \textit{5)} ordinary core-collapse supernovae (stars
over 10.3 $\Msun$).  Examples of each category are given in
\Tab{endstate}.  Compared with earlier similar studies, our mass
limits may be approximately $1\,\Msun$ lower than traditional values,
e.g., supernovae are often assumed to start at $8\,\Msun$, not
$7\,\Msun$.  While we do not place great faith in the exact values of
these masses, the existence of the various classes of events should be
robust for the one-dimensional stellar physics employed.

% fig 14 - outcomes
\begin{figure}
\includegraphics[width=\columnwidth]{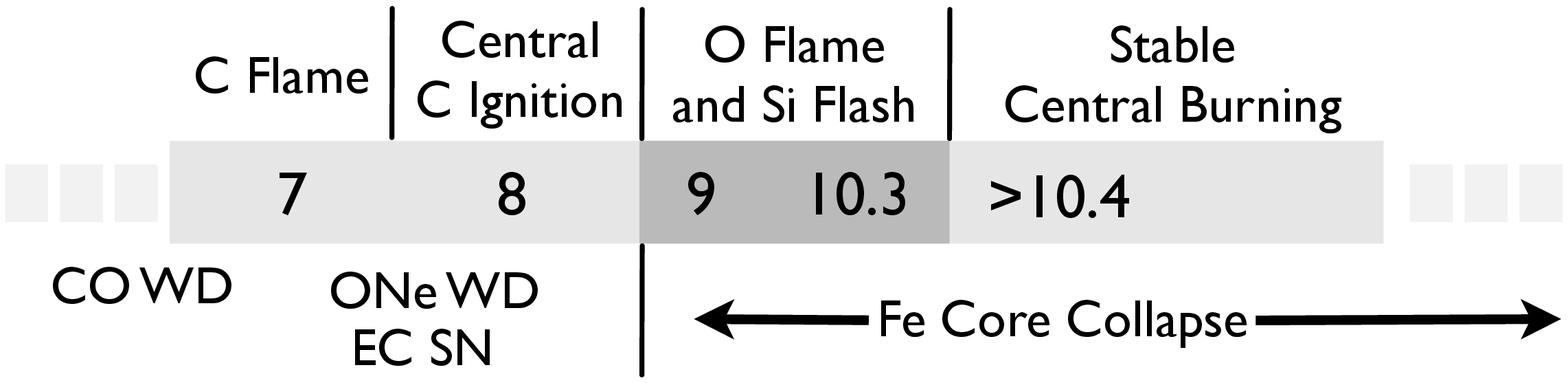}
\caption{The final fates of stars in the mass range 7 to 11 \Msun.
  Below $7\, \Msun$, CO white dwarfs are produced. From 7 to 8 \Msun,
  carbon ignites off center and burns as a CBF to the center and from
  8 to 9 carbon ignites centrally. From 7 to 9 \Msun, degenerate ONe
  cores are produced that may become white dwarfs if the envelope is
  lost, or electron-capture supernovae otherwise. Above 9.0 \Msun, all
  stars eventually produce iron cores that collapse to neutron
  stars. From 9.0 though 10.3 \Msun, silicon burning ignites in a
  strong flash that, especially in the more massive stars, can become
  a deflagration. Above 10.4 \Msun, all burning stages ignite in the
  center of the star without strong flashes. \lFig{fates}}
\end{figure}

We find, as have others, that those stars in this mass range
that do eventually produce neutron stars have compact structures.
that should be easy to explode using neutrinos, possibly too easy.
Without strong magnetic fields and rotation, which we argue are
negligible here, and without the ram pressure of an accreting mantle,
the resulting explosion will probably be much weaker than the
customary $10^{51}\,$erg inferred observationally for common
supernovae \citep[e.g.,][]{Kas09} and possibly more symmetric.

Though certainly not the last word, our treatment of CBFs in oxygen
and silicon burning is novel (\Sect{microflame}). Despite including
the relevant physics, including a large network and associated
neutrino losses, we do not find that thermohaline mixing or URCA
shells pay a major role in either stage. Flame propagation and
Rayleigh-Taylor instability, to the extent that either can be modeled
in a 1D calculation, dominate. The advancement of the burning is also
driven by macroscopic considerations (\Sect{macroflame}), the need to
provide fuel to maintain the entropy of a core whose mass already
exceeds the cold Chandrasekhar limit. Since the burning can only be
sustained by the inward propagation of a flame, the CBF cannot go out,
but must proceed, by whatever processes, at a rate at least sufficient
to balance neutrino losses from the core.

But we have also demonstrated the need for further work. The tables of
\citet{Tim94} for CBF speeds need to be extended, for oxygen-neon
compositions, to lower density, and similar tables calculated for
neutron-rich silicon burning. Perhaps more importantly, portions of
the oxygen- and silicon-CBF evolution need to be examined in 3D. This
could be done, for example, with an adaptive mesh, low-Mach-number
code like MAESTRO \citep{Non12}. Our 1D description of a CBF as a
discontinuous spherical shell is almost certainly wrong. In 3D, the
burning may be localized into numerous hot spots. Whether the
resulting temperature structure will be able to prevent a large scale
overturn of the composition inversion is a fascinating question.

A novel result of the present models \citep[though see][]{Woo80} is
the possibility of silicon igniting degenerately with a violent
flash. For the 9.2, 9.4, 9.5, and $9.6\, \Msun$ models, ordinary
convection was (barely) capable of transporting the energy from the
flash efficiently enough that no runaway on a hydrodynamical time
scale occurred. For the 9.0, 9.3 and 9.8 - $10.3\, \Msun$ models,
however, the flash was so powerful that ordinary convection was not
able to transport the energy and a localized deflagration developed.
In this regard, the silicon flash in these stars resembles a Type Ia
supernova in a Chandrasekhar carbon-oxygen white dwarf with an outcome
that is likely to be equally difficult to determine.  A key difference
is that, unlike carbon burning, silicon burning to iron-group nuclei
provides inadequate energy to unbind the core.  The rapid burning
does, however, cause an expansion of the core on a time scale that is
short compared with the hydrodynamic time of the surrounding helium
shell and base of the hydrogen envelope.

The observable outcome in these cases depends upon how much silicon
burns in the flash and, once again, 3D calculations will ultimately be
needed to answer the question. Here the efficiency of silicon
deflagration was parametrized by adjusting, artificially, a convective
efficiency parameter.  If the amount of silicon that burns is less
than about 0.3 \Msun, the silicon flash has little observable
consequence. Because of the small energy release, the core does not
expand greatly and recovers rapidly. The neutrino-mediated
Kelvin-Helmholtz time for the silicon-iron core is less than about two
weeks. If iron core collapse launches a second more powerful shock, as
seems likely, the shock from the silicon flash will be overtaken
either prior to or shortly after breakout. While the breakout
transient itself might be altered, the main light curve would not
change greatly. Unfortunately, based upon our understanding of similar
explosions in Type Ia supernovae, the amount burned in an off-center
ignited thermonuclear runaway may be small if no detonation occurs
\citep{Mal14}, and this may be the case for silicon deflagration as
well. If so, a rather ordinary Type IIp supernovae, will result,
albeit with lower than typical velocities and brightness on the
plateau (\Fig{10blite}).

On the other hand, if the amount of silicon that burns substantially
exceeds $0.3\,\Msun$, there will be an appreciable delay between envelope
ejection and core collapse (\Tab{outcome}). Burning more silicon gives
higher speed to the ejected envelope and lengthens the
Kelvin-Helmholtz time for the recontracting core. Maximum ejection
speeds for the envelope range from several hundred to several
thousand km$\,$s$^{-1}$.  For the more energetic explosions, those
that burned $0.7\,\Msun$ of silicon and more, the waiting time between
envelope ejection and final core collapse was months to years giving
rise to the possibility of two supernova-like events from the same
star.  The first event is far fainter than the second, and fainter
than typical Type IIp supernovae.

This behavior is intriguing given the historical record
for the Crab Supernova, SN~1054.  Because of the low mass of the
ejecta, the small abundances of heavy elements, and the presence of a
pulsar, this supernova has often been associated with the death of a
star that, on the main sequence, had a mass $\sim10\,\Msun$
\citep{Nom82,Wan09}.  Yet the historical record, such as it is,
suggests that the Crab was not unusually faint \citep{Smi13}.  We
agree with Smith that the Crab may have been brightened by a
substantial contribution from ``circumstellar interaction''.  To
appreciably alter the light curve at maximum, however, one needs a
substantial fraction of a solar mass of material at
$\sim$10$^{15}\,$cm.  The models we calculated here for stars in the
$9.8\,\Msun$ - $10.3\,\Msun$ range are capable of doing that provided
that the silicon flash is sufficiently violent to burn roughly half
the core.

A new calculation (\Sect{rotate}) of 10 \Msun \ evolution that
includes rotation and the effects of magnetic torques predicts a
rotation rate for the newly minted Crab pulsar of about 17 ms, which
compares quite favorably with the observationally inferred value at
birth, about 21 ms. The same calculation predicts a gravitational mass
from the Crab pulsar of from 1.2 to $1.3\, \Msun$.
%/Volumes/DATA1/q/qdat/woosley/stars/tensol2/10r/rerun
%BE/M \approx 0.6 \beta/(1 - 0.5 \beta)
%where $\beta = GM/(Rc^2)$.

Even if the silicon flash proves inadequate to eject the envelope
sufficiently in advance of core collapse, there is still the
possibility of substantial mass ejection by energy transported
acoustically from the vastly super-Eddington convective shells during
the last few years of the evolution
\citep[\Sect{wind};][]{Qua12,Shi14}.  Unlike supernovae in more massive
stars where the length of oxygen burning is a few months or less
\citep{Woo02}, oxygen shell burning igniting off-center and moving
inwards by a convectively bounded flame typically takes years to
reach the center. Even a small fraction of the integrated convective
luminosity during the last ten years would be sufficient to eject the
entire envelope.  The interaction of the supernova shock with a dense
wind could also appreciably brighten the event \citep{Mor14}.

\acknowledgements

We appreciate helpful correspondence with Adam Burrows and Thomas
Janka regarding the energy of models for accretion-induced collapse
and John Lattanzio regarding the evolution of AGB stars.  Tuguldur
Sukhbold helped greatly with preparing the figures, especially
\Fig{flames}.  This work was supported by NASA (NNX14AH34G) and the UC
Office of the President (12-LF-237070).  AH was support by an ARC
Future Fellowship (FT120100363).

\newpage

% Table 1
\begin{deluxetable}{ccccccccc}
\tabletypesize{\scriptsize}
\tablecaption{SUMMARY OF MODELS$^a$}
\tablehead{Initial & Final & Helium & CO & Si & Fe  & BE  & BE  & Outcome \\
Mass &  Mass &  Core Mass & Core Mass & Core Mass & Core Mass &  Envel & O-shell &   \\
(\Msun) & (\Msun) & (\Msun) & (\Msun) & (\Msun) & (\Msun)  &($-10^{47}\,$erg) & ($-10^{49}\,$erg) & }
\startdata
6.5     & 6.38 & 0.960 & 0.960 &   -   &   -   & 2.0$^b$  & -  &  CO  WD    \\
7.0     & 6.79 & 1.033 & 1.033 &   -   &   -   & 2.1$^b$  & -  &  OC  WD    \\
7.5     & 6.96 & 1.088 & 1.088 &   -   &   -   & 1.8$^b$  & - & ONe WD/EC SN \\
8.0     & 7.76 & 1.171 & 1.171 &   -   &   -   & 1.2$^b$ & -   &  ``        \\
8.5     & 8.28 & 1.271 & 1.271 &   -   &   -   & 2.3$^b$ & -   &   ``       \\
8.75    & 8.51 & 1.345 & 1.345 &   -   &   -   & 1.1$^b$ & -   &   ``       \\
9.0A$^c$& 8.75 & 1.567 & 1.400 & 1.320 & 1.237 &  -  & 5.8  & Si-Defl. SN   \\
9.0E$^c$& 8.75 & 1.567 & 1.400 & 1.320 & 1.268 &  -  &  -   & Si-Defl. SN    \\
9.1     & 8.83 & 1.640 & 1.418 & 1.334 & 1.244 & 1.8 & 4.8  & Si-Flash SN  \\
9.2     & 8.93 & 1.759 & 1.441 & 1.365 & 1.244 & 2.0 & 3.4  & Si-Flash SN  \\
9.3A$^c$& 9.02 & 1.856 & 1.457 & 1.346 & 1.280 & -   & 8.0  & Si-Defl. SN  \\
9.3E$^c$& 9.02 & 1.856 & 1.438 & 1.363 & 1.261 & -   &  -   & Si-Defl. SN  \\
9.4     & 9.11 & 1.975 & 1.477 & 1.397 & 1.331 & 2.2 & 4.1  & Si-Flash SN   \\
9.5     & 9.21 & 2.054 & 1.493 & 1.356 & 1.332 & 2.2 & 6.8  & Si-Flash SN   \\
9.6     & 9.30 & 2.094 & 1.528 & 1.400 & 1.302 & 2.1 & 7.0 &    ``         \\
9.7     & 9.39 & 2.183 & 1.546 & 1.412 & 1.305 & 2.2 & 7.4 &    ``         \\
9.8A$^c$ & 9.48 & 2.281 & 1.564 & 1.409 & 1.316 & 0.3 & 9.4 &  Si Defl. SN \\
9.8E$^c$ & 9.48 & 2.281 & 1.564 & 1.269 & 1.215 &  -  & 9.0 &       ``     \\
9.9A$^c$ & 9.58 & 2.356 & 1.588 & 1.415 & 1.349 & 1.0 & 10  &      ``     \\
9.9E$^c$ & 9.58 & 2.356 & 1.597 & 1.302 & 1.231 &  -  & -  &       ``     \\
10.0A$^c$& 9.69 & 2.448 & 1.612 & 1.430 & 1.362 & 1.5 & 11  &       ``     \\
10.0E$^c$& 9.69 & 2.448 & 1.626 & 1.311 & 1.232 &  -  & -   &       ``     \\
10.1C$^c$& 9.79 & 2.484 & 1.634 & 1.427 & 1.354 &  -  & 9.9  &       ``     \\
10.1E$^c$& 9.79 & 2.484 & 1.657 & 1.336 & 1.256 &  -  &  -  &       ``     \\
10.2C$^c$& 9.89 & 2.545 & 1.655 & 1.427 & 1.363 &  -  & 10  &       ``     \\
10.2E$^c$& 9.89 & 2.545 & 1.638 & 1.370 & 1.296 &  -  & -   &       ``     \\
10.3D$^c$& 10.00& 2.591 & 1.670 & 1.438 & 1.336 &  -  & 8.3 &       ``     \\
10.3E$^c$& 10.00& 2.591 & 1.645 & 1.363 & 1.260 &  -  &  -  &       ``     \\
10.4   & 10.09 & 2.634 & 1.684 & 1.477 & 1.353 & 2.0 & 9.9 &  Ordinary SN \\
10.5   & 10.19 & 2.666 & 1.709 & 1.477 & 1.355 & 2.0 & 11  &       ``     \\
10.75  & 10.45 & 2.736 & 1.736 & 1.523 & 1.318 & 2.5 & 7.5 &       ``     \\
11     & 10.68 & 2.797 & 1.780 & 1.545 & 1.411 & 2.8 & 7.3 &       ``     \\
11.25  & 10.90 & 2.835 & 1.802 & 1.552 & 1.346 & 2.8 & 8.1 &       ``     \\
11.5   & 10.81 & 2.740 & 1.757 & 1.487 & 1.375 & 2.7 & 12.5 &      ``     \\
12.0   & 10.93 & 3.103 & 1.997 & 1.636 & 1.290 & 3.7 & 12.2 &      ``     \\
12.25  & 11.08 & 3.198 & 2.071 & 1.546 & 1.445 & 4.6 & 23.1 &      ``     \\
12.5   & 11.23 & 3.304 & 2.145 & 1.567 & 1.464 & 4.8 & 24.0 &      ``     \\
12.75  & 11.39 & 3.400 & 2.216 & 1.582 & 1.472 & 5.7 & 25.5 &      ``     \\
13.0   & 11.59 & 3.486 & 2.282 & 1.602 & 1.489 & 6.3 & 27.3 &      ``     \\
13.25  & 11.68 & 3.607 & 2.377 & 1.644 & 1.512 & 7.3 & 29.7 &      ``     \\
13.5   & 11.93 & 3.675 & 2.434 & 1.658 & 1.518 & 7.9 & 30.9 &      ``     \\
\enddata
\tablenotetext{a}{For stars below $9.0\,\Msun$, the star was not
  evolved to a presupernova. Data is given for the last model
  calculated. For other models the values given are either for the
  presupernova star or at the silicon flash.}
\tablenotetext{b}{Because of poor zoning at edge of the edge of the
  degenerate core, binding energies for models less than $9.0\,\Msun$
  are not very accurate.}  
\tablenotetext{c}{For stars with $9.0\,\Msun$, $9.3\, \Msun$, and
  $9.8\,\Msun$ to $10.3\,\Msun$, the binding energies outside of the
  fiducial cores may become positive and are indicated with a
  ``-''. The CO and Si core masses vary with the strength of the
  silicon flash and results from both a weak and strong flash are
  given (see also \Tab{outcome}).}  
\lTab{endstate}
\end{deluxetable}

\clearpage

%Table 2
\begin{deluxetable}{cccccccc}
\tablecaption{SILICON IGNITION CONDITIONS}
\tablehead{Initial & Silicon    & Ignition &     &     &   &    &  \\
              Mass & Core Mass$^a$ & Mass     & Density &
Temperature & $\eta$ & Luminosity$^b$  & Radius$^a$ \\
 (\Msun) & (\Msun) &  (\Msun) & (g cm$^{-3}$)  & (K) &
&($10^{38}\,$erg$\,$s$^{-1}$) & ($10^{13}\,$cm)  }
\startdata
9.0   & 1.308  & 0.408  & 4.94(8)  & 3.21(9) & 9.43 & 1.04 & 2.83  \\
9.1   & 1.334  & 0.358  & 4.94(8)  & 3.25(9) & 9.30 & 1.02 & 2.78  \\
9.2   & 1.365  & 0.314  & 4.58(8)  & 3.21(9) & 9.16 & 1.00 & 2.75  \\
9.3   & 1.343  & 0.330  & 4.60(8)  & 3.20(9) & 9.21 & 1.01 & 2.73  \\
9.4   & 1.387  & 0.275  & 4.15(8)  & 3.15(9) & 8.96 & 1.04 & 2.77  \\
9.5   & 1.332  & 0.237  & 4.67(8)  & 3.19(9) & 9.31 & 1.08 & 2.83  \\
9.6    & 1.399 & 0.2002 & 4.87(8)  & 3.20(9) & 9.39 & 1.12 & 2.96 \\
9.7    & 1.412 & 0.1527 & 4.84(8)  & 3.20(9) & 9.39 & 1.15 & 3.02 \\
9.8    & 1.409 & 0.0582 & 5.22(8)  & 3.20(9) & 9.67 & 1.20 & 3.05 \\
9.9    & 1.415 & 0.0364 & 5.17(8)  & 3.20(9) & 9.64 & 1.26 & 3.11 \\
10.0   & 1.427 & 0.0203 & 4.97(8)  & 3.23(9) & 9.40 & 1.35 & 3.28 \\
10.1   & 1.427 & 0.0164 & 4.62(8)  & 3.16(9) & 9.35 & 1.39 & 3.34 \\
10.2   & 1.432 & 0.0020 & 4.53(8)  & 3.16(9) & 9.28 & 1.45 & 3.42 \\
10.3   & 1.438 &    0   & 4.55(8)  & 3.12(9) & 9.33 & 1.51 & 3.50 \\
10.4   & 1.441 &    0   & 4.03(8)  & 3.04(9) & 9.22 & 1.59 & 3.71 \\
10.5   & 1.370 &    0   & 1.90(8)  & 3.10(9) & 8.87 & 1.63 & 3.68 \\
10.75  & 1.421 &    0   & 3.34(8)  & 3.10(9) & 8.34 & 1.73 & 3.78 \\
11.0   & 1.411 &    0   & 3.14(8)  & 3.10(9) & 8.11 & 1.80 & 3.87 \\
11.25  & 1.404 &    0   & 3.08(8)  & 3.10(9) & 8.06 & 1.85 & 3.91 \\
11.5   & 1.401 &    0   & 3.50(8)  & 3.10(9) & 8.49 & 1.70 & 3.72 \\
11.75  & 1.409 &    0   & 2.92(8)  & 3.10(9) & 7.86 & 2.03 & 4.17 \\
12.0   & 1.451 &    0   & 2.36(8)  & 3.10(9) & 7.20 & 2.16 & 4.32 \\
12.25  & 1.550 &    0   & 2.47(8)  & 3.10(9) & 7.32 & 2.28 & 4.48 \\
12.5   & 1.565 &    0   & 2.26(8)  & 3.10(9) & 7.07 & 2.36 & 4.58 \\
12.75  & 1.580 &    0   & 2.24(8)  & 3.10(9) & 7.05 & 2.49 & 4.71 \\
13.0   & 1.600 &    0   & 2.07(8)  & 3.10(9) & 6.81 & 2.58 & 4.73 \\
13.25  & 1.642 &    0   & 1.86(8)  & 3.10(9) & 6.50 & 2.71 & 4.88 \\
13.5   & 1.656 &    0   & 1.74(8)  & 3.10(9) & 6.32 & 2.80 & 4.94 \\
\enddata
\tablenotetext{a}{The mass of the oxygen depleted core at the time the
  silicon flash occurred. Not necessarily the silicon core mass of the
  presupernova star.}
\tablenotetext{b}{Evaluated at central oxygen depletion.}
\lTab{siflash}
\end{deluxetable}

\begin{deluxetable}{cccccccc}
\tablecaption{SUMMARY SILICON DEFLAGRATIONS}
\tablehead{Initial & Fe   & Conv.  & Delay & Photo. & Edge & Velocity & Lum.\\
              Mass & Mass$^a$ &  param & Time  & Radius & PreSN & Edge    & PreSN \\
 (\Msun) & (\Msun) &  & (sec)  & (cm) &  & (cm) & (erg$\,$s$^{-1}$) }
\startdata
9.0A & 0.28  & 1.(-4) & 8.5(5)  & 2.8(13) & 2.8(13) &   0    & 1.0(38) \\
9.0E & 0.56  & 0.01   & 1.14(7) & 3.3(14) & 2.3(15) & 3.6(8) & 7.2(40) \\
9.3A & 0.33  & 1.(-4) & 8.7(5)  & 2.7(13) & 2.7(13) &   0    & 1.0(38) \\
9.3E & 0.59  & 0.01   & 8.9(6)  & 3.6(14) & 2.1(15) & 3.1(8) & 7.4(40) \\
9.8A  & 0.21 & 1.(-4) & 1.33(6) & 3.0(13) & 3.1(13) & 0      & 1.2(38) \\
9.8B  & 0.29 & 3.(-4) & 1.39(6) & 3.8(13) & 3.8(13) & 4.1(7) & 1.9(39) \\
9.8C  & 0.63 & 6.(-4) & 1.09(7) & 3.2(14) & 1.8(15) & 1.8(8) & 8.4(40) \\
9.8D  & 0.76 & 0.001  & 4.66(7) & 1.1(13) & 9.5(15) & 2.1(8) & 7.1(39) \\
9.8E  & 0.77 & 0.01   & 5.97(7) & 1.1(11) & 1.9(16) & 3.6(8) & 4.9(39) \\
9.9A  & 0.22 & 1.(-4) & 1.34(6) & 3.1(13) & 3.1(13) & 0      & 1.3(38) \\
9.9B  & 0.28 & 3.e-4) & 1.39(6) & 3.5(13) & 2.5(13) & 4.3(7) & 4.2(39) \\
9.9C  & 0.36 & 6.(-4) & 2.25(6) & 1.2(14) & 1.5(14) & 7.9(7) & 1.3(40) \\
9.9D  & 0.76 & .001   & 4.08(7) & 3.9(13) & 9.0(15) & 2.2(8) & 7.0(40) \\
9.9E  & 0.78 & 0.01   & 5.22(7) & 3.8(13) & 1.3(16) & 3.3(8) & 1.3(40) \\
10.0A & 0.18 & 1.(-4) & 1.33(6) & 3.3(13) & 3.3(13) &  0     & 1.4(38) \\
10.0B & 0.25 & 3.(-4) & 1.38(6) & 3.3(13) & 3.3(13) &  0     & 1.4(38) \\
10.0C & 0.34 & 6.(-4) & 1.95(6) & 9.3(13) & 1.1(14) & 7.7(7) & 1.0(40) \\
10.0C+ & 0.61 & 9.e-4  & 8.87(6) & 3.6(14) & 1.5(15) & 1.9(8) & 1.3(41) \\
10.0D & 0.76 & 0.001  & 3.46(7) & 3.7(13) & 8.3(15) & 2.5(8) & 2.7(40) \\
10.0E & 0.79 & 0.01   & 4.85(7) & 8.7(10) & 1.6(16) & 3.6(8) & 4.9(39) \\
10.1C & 0.32 & 6.(-4) & 1.72(6) & 6.8(13) & 7.3(13) & 6.0(7) & 6.3(39) \\
10.1D & 0.39 & 0.001  & 2.60(6) & 1.5(14) & 2.1(14) & 9.5(7) & 2.1(40) \\
10.1E & 0.77 & 0.01   & 3.42(7) & 5.0(13) & 8.0(15) & 2.8(8) & 1.2(40) \\
10.2C & 0.30 & 6.(-4) & 1.48(6) & 4.5(13) & 4.5(13) & 5.2(7) & 5.0(39) \\
10.2D & 0.71 & 0.001  & 1.48(7) & 2.1(13) & 3.4(15) & 2.6(8) & 1.3(39) \\
10.2E & 0.73 & 0.01   & 1.90(7) & 4.2(13) & 4.7(15) & 2.8(8) & 5.8(39) \\
10.3D & 0.36 & 0.001  & 2.23(6) & 1.2(14) & 1.4(14) & 8.0(7) & 1.4(40) \\
10.3E & 0.74 & 0.01   & 2.13(7) & 3.4(13) & 5.0(15) & 2.6(8) & 3.6(39) \\
\enddata
\tablenotetext{a}{The amount of intermediate mass elements that fused
  to iron-group elements during the flash.}
\lTab{outcome}
\end{deluxetable}

\begin{deluxetable}{cccccccccc}
\tablecaption{Nucleosynthesis for Models~10B and 10D (He Core only)}
\tablehead{Model   & Piston   & Energy         & O & Mg  & Si & S & Ar & Ca &
$^{56}$Ni \\
                   & Location & ($10^{50}\,$erg) &(\Msun) & (\Msun) & (\Msun)
 & (\Msun)  & (\Msun)  & (\Msun)  & (\Msun) }
\startdata
10.0B & 1.301 & 2.0 & 0.161 & 0.010 & 0.048 & 0.034 & 0.0065 & 0.0015 & 0.0060 \\
10.0B & 1.245 & 1.8 & 0.161 & 0.010 & 0.053 & 0.038 & 0.0073 & 0.0016 & 0.045  \\
10.0D & 1.440 & 2.2 & 0.099 & 0.0060 & 0.011 & 0.0042 & 0.0008 & 0.0007 & 0.022 \\
\enddata
\lTab{nucleo}
\end{deluxetable}

\end{document}